\documentclass[reprint,superscriptaddress,amsmath,amssymb,aps,prx]{revtex4-2}
\usepackage{graphicx}
\usepackage{dcolumn}
\usepackage{bm}
\usepackage{qcircuit}
\usepackage{algpseudocode}
\usepackage{graphicx}
\usepackage{algorithm}
\usepackage{multirow}
\usepackage{booktabs}
\usepackage{subfigure}
\usepackage[pass]{geometry}
\usepackage{float}
\usepackage{amsthm}
\usepackage{physics}
\usepackage[version=4]{mhchem}
\usepackage{diagbox}
\usepackage{xcolor}
\usepackage{bibunits}
\usepackage[justification=raggedright]{caption}
\makeatletter
\renewcommand*\env@matrix[1][\arraystretch]{%
\edef\arraystretch{#1}%
\hskip -\arraycolsep
\let\@ifnextchar\new@ifnextchar
\array{*\c@MaxMatrixCols c}}
\makeatother
\begin{document}
\preprint{APS/123-QED}
\title{Taming Rydberg Decay with Measurement-based Quantum Computation}
\author{Cheng-Cheng Yu}
\affiliation{Hefei National Research Center for Physical Sciences at the Microscale and School of Physical Sciences, University of Science and Technology of China, Hefei 230026, China}
\affiliation{Shanghai Research Center for Quantum Science and CAS Center for Excellence in Quantum Information and Quantum Physics,
University of Science and Technology of China, Shanghai 201315, China}
\affiliation{Hefei National Laboratory, University of Science and Technology of China, Hefei 230088, China}
\author{Zi-Han Chen}
\affiliation{Hefei National Research Center for Physical Sciences at the Microscale and School of Physical Sciences, University of Science and Technology of China, Hefei 230026, China}
\affiliation{Shanghai Research Center for Quantum Science and CAS Center for Excellence in Quantum Information and Quantum Physics,
University of Science and Technology of China, Shanghai 201315, China}
\affiliation{Hefei National Laboratory, University of Science and Technology of China, Hefei 230088, China}
\author{Yu-Hao Deng}
\email{dengyh@ustc.edu.cn}
\affiliation{Hefei National Research Center for Physical Sciences at the Microscale and School of Physical Sciences, University of Science and Technology of China, Hefei 230026, China}
\affiliation{Shanghai Research Center for Quantum Science and CAS Center for Excellence in Quantum Information and Quantum Physics,
University of Science and Technology of China, Shanghai 201315, China}
\affiliation{Hefei National Laboratory, University of Science and Technology of China, Hefei 230088, China}
\author{Chao-Yang Lu}
\email{cylu@ustc.edu.cn}
\affiliation{Hefei National Research Center for Physical Sciences at the Microscale and School of Physical Sciences, University of Science and Technology of China, Hefei 230026, China}
\affiliation{Shanghai Research Center for Quantum Science and CAS Center for Excellence in Quantum Information and Quantum Physics,
University of Science and Technology of China, Shanghai 201315, China}
\affiliation{Hefei National Laboratory, University of Science and Technology of China, Hefei 230088, China}
\affiliation{New Cornerstone Science Laboratory, Hefei, 230026, China}
\author{Ming-Cheng Chen}
\email{cmc@ustc.edu.cn}
\affiliation{Hefei National Research Center for Physical Sciences at the Microscale and School of Physical Sciences, University of Science and Technology of China, Hefei 230026, China}
\affiliation{Shanghai Research Center for Quantum Science and CAS Center for Excellence in Quantum Information and Quantum Physics,
University of Science and Technology of China, Shanghai 201315, China}
\affiliation{Hefei National Laboratory, University of Science and Technology of China, Hefei 230088, China}

\author{Jian-Wei Pan}
\email{pan@ustc.edu.cn}
\affiliation{Hefei National Research Center for Physical Sciences at the Microscale and School of Physical Sciences, University of Science and Technology of China, Hefei 230026, China}
\affiliation{Shanghai Research Center for Quantum Science and CAS Center for Excellence in Quantum Information and Quantum Physics,
University of Science and Technology of China, Shanghai 201315, China}
\affiliation{Hefei National Laboratory, University of Science and Technology of China, Hefei 230088, China}
\begin{abstract}
Programmable neutral atom arrays show great promise for fault-tolerant quantum computing. A dominant physical error on this platform is qubit leakage and loss, notably decay errors from the Rydberg state during two-qubit gates. Such leakage events are particularly detrimental as they propagate, generating correlated errors that severely degrade the effective error distance of quantum error correction codes. Here, we present a novel approach to address Rydberg decay errors leveraging measurement-based quantum computation (MBQC). Our scheme strategically exploits the inherent geometric structure of topological cluster states and only uses final leakage detection information to locate propagated errors originating from Rydberg decay. This eliminates the need for complex and atom-species-specific mid-circuit leakage detection, offering broader applicability, e.g., to the well-established Rb atom platform. We demonstrate a high error threshold of 3.65\% per CZ gate for pure Rydberg decay and achieve a favorable error distance $d_e \approx d$. Our method compares favorably with state-of-the-art erasure conversion protocols in the sub-threshold performance, offering comparable or marginally larger logical error rates while significantly reducing experimental overhead.
\end{abstract}
\maketitle
Neutral atom arrays have emerged as a promising platform for quantum computing, owing to their high-fidelity two-qubit quantum operations \cite{evered2023high,scholl2023erasure,ma2023high,PhysRevX.15.011009}, scalability \cite{bluvstein2022quantum,bluvstein2024logical}, and continuous loading \cite{v7ny-fg31,Chiu_2025,li2025fastcontinuouscoherentatom}. For neutral atoms, the leakage error from the Rydberg state (we refer to this as the Rydberg decay error or Rydberg decay), including blackbody radiation (BBR) and radiative decay (RD), is a dominant error source \cite{wu2022erasure,cong2022hardware,PRXQuantum.5.040343}. Leakage error is particularly detrimental for quantum error correction \cite{Dennis_2002,Bluvstein_2025,aliferis2007fault} because a single leakage error may induce a two-qubit error chain through multi-qubit gates, which leads to a degraded error distance $\lfloor \frac{d+3}{4} \rfloor$ compared to Pauli error \cite{PhysRevA.88.042308,suchara2015leakage,brown2020critical,jandura2024surfacecodestabilizermeasurements}. Researchers have shown that with mid-circuit leakage detection, the Rydberg decay error can be converted to an erasure error, which has an error distance $d_e = d$ \cite{wu2022erasure,kang2023quantum,ma2023high,scholl2023erasure,sahay2023high}. However, mid-circuit leakage detection is limited to certain species of atoms. For $\ce{^{171}_{}Yb}$ atoms, BBR and RD can be detected by the fast ionization and the cyclic transition, respectively \cite{wu2022erasure,ma2023high,omanakuttan2024coherencepreservingleakagedetection}. Some preliminary demonstrations have shown that decay errors to ground state in $\ce{^{171}_{}Yb}$ and $\ce{^{88}_{}Sr}$ are converted to erasure error \cite{ma2023high,scholl2023erasure}. But for alkali atoms, the detection of Rydberg decay requires measurement of an auxiliary qubit, which leads to significant time overhead if mid-circuit leakage detection is needed \cite{cong2022hardware,PRXQuantum.5.040343}.

An equivalent model for quantum computation is measurement-based quantum computation (MBQC), in which large-scale entangled states are prepared in advance and measured sequentially to implement quantum operations and process information \cite{PhysRevLett.86.5188,PhysRevA.68.022312,briegel2009measurement}. Fault-tolerance of MBQC can be achieved by 3D Raussendorf-Harrington-Goyal (RHG) cluster states \cite{raussendorf2006fault,raussendorf2007topological,PhysRevLett.98.190504,Tournaire_2026} that can be mapped to 2D surface codes propagating in time \cite{PhysRevLett.117.070501,PhysRevResearch.2.033305,Baranes_2026}. Importantly, leakage errors, a dominant error source in neutral atoms, can be naturally addressed in MBQC \cite{PhysRevA.90.052316,aliferis2007fault}, where all qubits have a constant lifetime. In circuit-based error correction (such as the surface code), only a subset of qubits are measured when implementing syndrome measurement. Therefore, if a leakage error occurs on the unmeasured qubit, it remains undetected unless we introduce additional leakage detection \cite{PhysRevA.90.052316,suchara2015leakage}. However, in MBQC, every qubit is frequently measured, so the leakage can be detected to convert the error to a less harmful erasure error \cite{PhysRevA.90.052316,aliferis2007fault}.

In this article, we propose a novel strategy to address Rydberg decay with MBQC. First, we reveal that when the leaked state of Rydberg decay $\ket{L}$ is not involved in subsequent single- and two-qubit quantum operations, the resulting error channel has a similar error propagation pattern to Pauli errors and thus does not degrade the distance. Building on this, we take advantage of the \textit{located error} \footnote{We select this terminology to distinguish from \textit{erasure error}, as previously it is generally used to refer to \textit{detected leakage}. But one essence of our work is that we use indirect final measurement information to locate the error, so it is different from the traditional erasure error.} to address the propagated error from a decoding perspective. Specifically, we utilize the inherent structure of the RHG cluster state to \textit{locate} the propagated error, with a final three-outcome measurement to distinguish leaked state ($\ket{L}$) and states in qubit subspace ($\ket{0},\ket{1}$)\cite{suchara2015leakage,PRXQuantum.5.040343,Baranes_2026}. The whole protocol is realized with no need for mid-circuit leakage detection, thereby offering substantial hardware overhead reduction and broader applicability, e.g., to the well-established Rb atom platform.

Notably, the reduced experimental requirements don't deteriorate our performance drastically. Instead, we demonstrate a high threshold 3.65\% for each CZ gate and an error distance $d_e \approx d$, for pure Rydberg decay. Regarding a mixture of Rydberg decay and two-qubit depolarization errors in CZ gates, we demonstrate comparable performance to the biased erasure conversion \cite{sahay2023high}, at low physical error rates. Therefore, our work paves an experiment-friendly way for neutral atom qubits to achieve high performance in the presence of Rydberg decay, and suggests that MBQC holds significant potential for applications involving neutral atoms.

\begin{figure}[!htbp]
    \centering
    \includegraphics[width=0.25\textwidth]{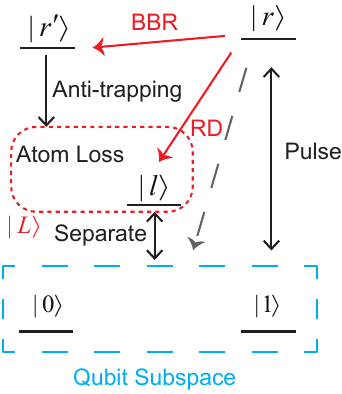}
    \captionsetup{justification=raggedright}
    \caption{Energy levels and Rydberg decay in neutral atoms: Two low-lying states encode the qubit and only the state $\ket{1}$ is coupled to $\ket{r}$. The BBR branch takes the atom to other Rydberg states and is converted to atom loss with an anti-trapping potential or a microwave-assisted process. The majority of the RD branch transitions the atoms to some other low-lying states $\ket{l}$, which are separate from the qubit subspace. The leaked state $\ket{L}$ represents both atom loss from BBR and other low-lying states $\ket{l}$ from RD.}
    \label{fig1}
\end{figure}

\textit{Physical error model and Pauli twirling.} For a frequently-used encoding for computation, we use two low-lying levels to encode the qubit, with only the state $\ket{1}$ coupled to the Rydberg state (Fig.\ref{fig1}). The channel of Rydberg decay is described in the operator-sum form as $\xi(\rho) = \sum_{i=0,1} K_i \rho K_i^{\dagger}$ \cite{sahay2023high,cong2022hardware} with the Kraus operators below ($p_e$ is the Rydberg decay error rate).
\begin{equation}
\begin{cases}
K_0 = \ket{0}\bra{0} + \sqrt{1-p_e}\ket{1}\bra{1} + \ket{L}\bra{L}\\
K_1 = \sqrt{p_e}\ket{L}\bra{1}
\end{cases}
\label{eqn1}
\end{equation}
This channel becomes \textit{biased erasure} error under Pauli twirling, if we add leakage detection after each two-qubit gate and replace each leaked atom by a new atom in $\ket{1}$ \cite{sahay2023high}. Without atom renewal, the gate action on leaked atoms and the efficacy of Pauli twirling against coherent errors become ill-defined \cite{Bravyi_2018,Wallman_2016}. We prove that when the physical gate (single- and two-qubit gate) does not interact with the leaked qubit $\ket{L}$, the Pauli-twirling still applies to remove the coherent error but turned the \textit{biased erasure} to \textit{erasure} channel, namely previous evolution governed by jump operator $K_1$ is now governed by $K_{0L} = \sqrt{p_e/2} \ket{L}\bra{0}$ and $K_{1L}=\sqrt{p_e/2} \ket{L}\bra{1}$. Additionally, the leakage error (jump operator $K_{0L}$ or $K_{1L}$) exhibits a favorable error propagation property: it propagates similarly to a Pauli error. This is a key factor contributing to the improved error distance observed in our work \cite{PhysRevA.88.042308,suchara2015leakage,brown2020critical,jandura2024surfacecodestabilizermeasurements,d1v7-nctj,supp}.\\

\begin{figure*}[!htbp]
    \centering
    \includegraphics[width=\textwidth]{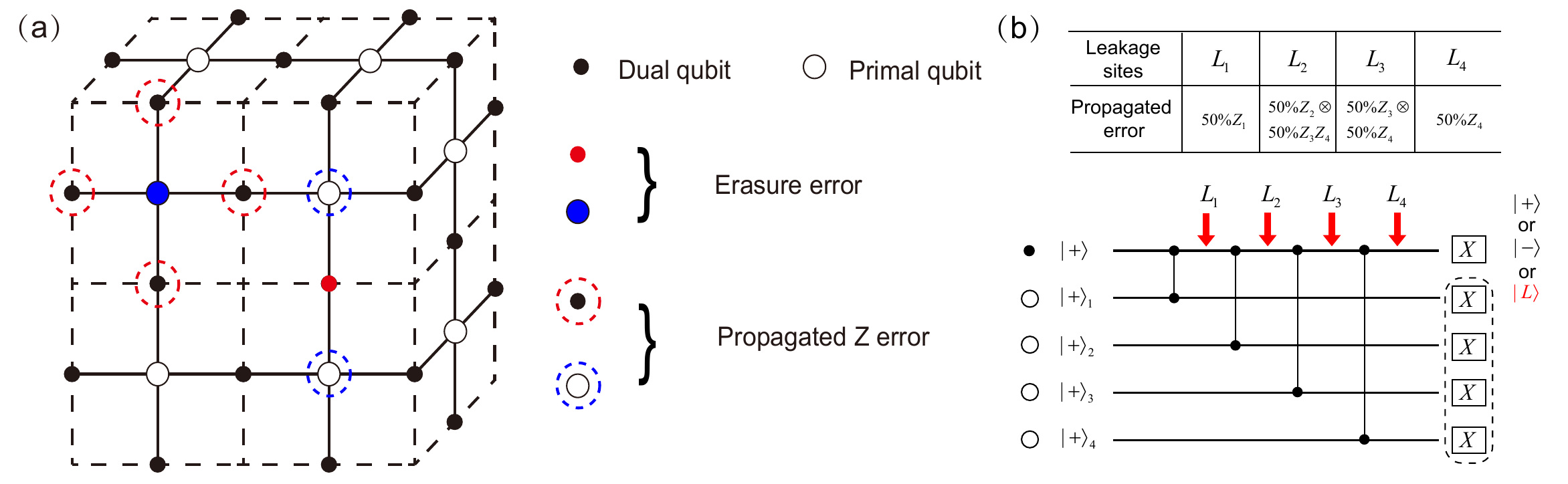}
    \captionsetup{justification=raggedright}
    \caption{The decoding protocol for Rydberg decay in the RHG cluster state \textbf{(a)} a $2\times2\times1$ lattice structure of the RHG cluster state. In a cell of the RHG cluster state, each dual qubit is connected to four primal qubits via CZ gates, and similarly, each primal qubit is connected to four dual qubits. For decoding the primal lattice, we need to account for erasure errors or directly detected leakage in primal qubits and propagated errors from dual qubits. If a leakage is detected in a dual qubit, we can calculate the propagated error probabilities with the circuit shown in (b). For decoding the dual lattice, the same procedure is implemented by exchanging the roles of the primary qubit and the dual qubit. \textbf{(b)} A (partial) circuit to generate the RHG cluster state when we consider propagated Z errors in primal qubits. Each qubit is initialized in $\ket{+}$ and is connected to neighboring qubits by a CZ gate. If a leakage is finally detected in a dual qubit, it happens at four error sites with approximately equal probability 25\% ($L_i$ represents that the dual qubit leaks at the \textit{i-th} CZ gate). Each error site associated with dual qubits triggers specific error mechanisms, represented as edges in the detector error model \cite{Gidney_2021}, as explicitly listed.}
    \label{fig2}
\end{figure*}

\textit{RHG cluster state, final three-outcome measurement, and located decoder.} Here, we consider a specific topological cluster state, the RHG cluster state, to implement MBQC. It can be regarded as a toric code or a surface code propagating in time. The structure of the RHG cluster state is shown in Fig.\ref{fig2}(a). To prepare the RHG cluster state, each qubit is initialized to $\ket{+}$, and we apply CZ gates between the connected vertices. A depth-4 circuit to generate the state is shown in Fig.\ref{fig2}(b). After state generation, layers of qubits are measured successively to process the logical information \cite{supp}.

Our method requires \textit{final three-outcome measurement}, through which we distinguish among the states $\ket{0}$, $\ket{1}$ and $\ket{L}$. Unlike current mid-circuit leakage detection, which is costly due to the use of extra ancillas for extracting leakage information without destroying data qubits \cite{cong2022hardware,PRXQuantum.5.040343} or is limited to certain atom species \cite{wu2022erasure,ma2023high}, the leakage/loss-resolving projective measurement is easier and applicable to general neutral atom species. For instance, the leakage/loss-resolving measurement on $\ce{^{87}_{}Rb}$ atoms is feasible with existing techniques \cite{PhysRevApplied.19.034089,Bluvstein_2025,supp}.

\begin{figure}[!htbp]
    \centering
    \includegraphics[width=0.45\textwidth]{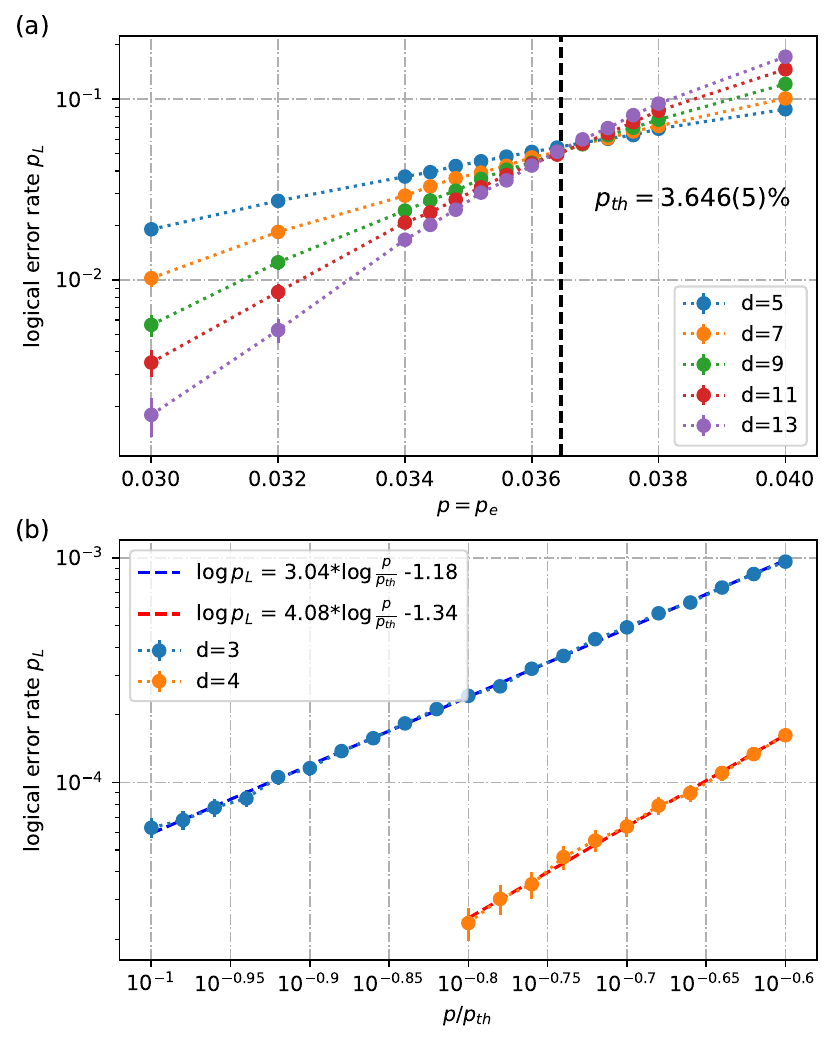}
    \captionsetup{justification=raggedright}
    \caption{Performance of pure Rydberg decay. \textbf{(a)} The threshold, \textbf{(b)} The error distances. It shows that Rydberg decay has a high threshold approximating $3.646\%$ and an error distance $d_e \approx d$ with a located decoder.}
    \label{fig3}
\end{figure}

\begin{figure}[!htbp]
    \centering
    \includegraphics[width=0.48\textwidth]{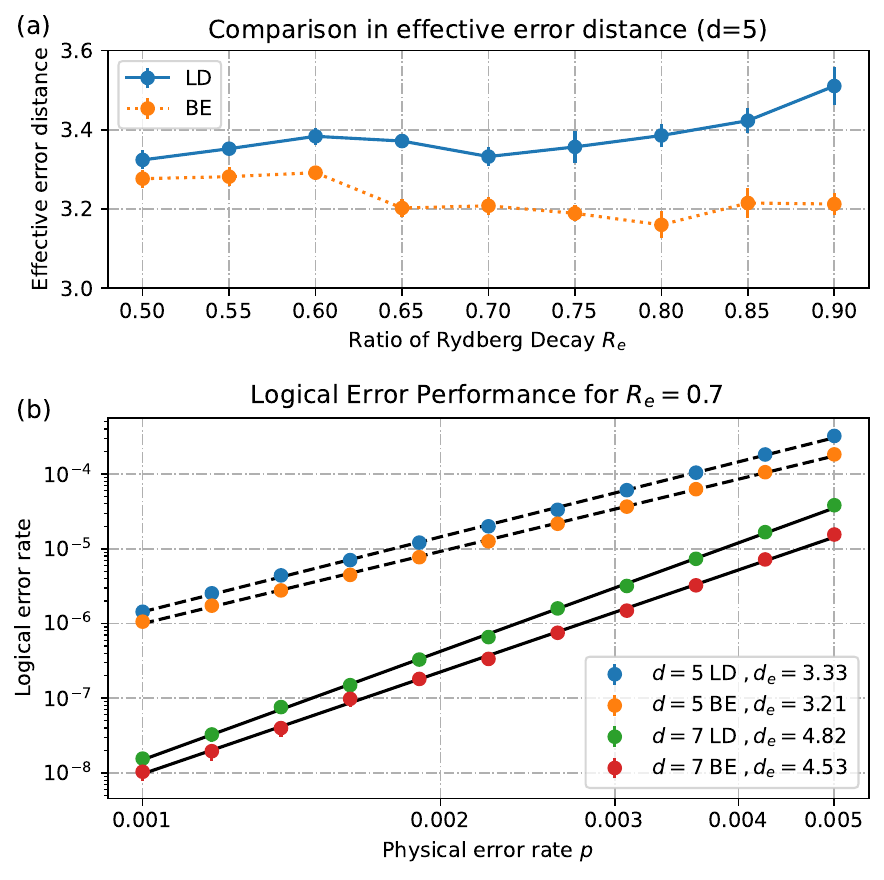}
    \captionsetup{justification=raggedright}
    \caption{Comparison between the located decoder (LD) and biased erasure conversion (BE).\textbf{(a)} Comparison in effective error distance, derived from linear-regression on $\log p_L$ and $\log p$, with physical error rate uniformly-spaced in $0.1\%$ to $0.5\%$ in logscale.\textbf{(b)} Logical error performance for $R_e = 0.7$. The advantage in the error distance persists for $d=7$, and the gap between these two schemes remains small. $\eta = \frac{1.56\times10^{-8}}{1.04\times10^{-8}} = 1.5$ for $d=7$ and $p=0.1\%$.}
    \label{fig4}
\end{figure}

Identifying atoms in $\ket{L}$ allows us to exploit \textit{located errors} to enhance decoding performance. \textit{Located error} means that we know which qubit or gate is susceptible to being noisy. A quantum code with distance $d$ corrects $\frac{d-1}{2}$ Pauli errors, but corrects $d-1$ \textit{located errors}, given proper decoders \cite{Nielsen_Chuang_2010}. So an outstanding feature of the \textit{located error} is a better error distance $d_e = d$ compared to the Pauli error $d_e=\frac{d+1}{2}$, which leads to faster suppression of the logical error rate in the sub-threshold regime ($p_L\sim (p/p_{th})^{d_e}$) \cite{wu2022erasure,985g-58gd}. Besides, it also has a significantly higher threshold compared to Pauli error, as it is easier to correct \cite{wu2022erasure,sahay2023high,985g-58gd}. While prior work has mainly concentrated on the characteristics of erasure errors (i.e., located errors with precisely known locations) \cite{wu2022erasure,sahay2023high,985g-58gd,d1v7-nctj,PhysRevResearch.7.013249}, our study demonstrates that located errors determined only by final measurements can retain much of the benefits in error correction performance \cite{985g-58gd,d1v7-nctj}. These are illustrated in the subsequent results in Fig.\ref{fig3} and Fig.\ref{fig4}.

We demonstrate how the final leakage detection is used to locate Rydberg decay errors. In the RHG cluster state, the primal lattice and the dual lattice are decoded separately, and we take decoding the primal lattice as an example. The detected leakage in primal qubits directly becomes \textit{erasure error}, which means that the primal qubit encounters a $50\%$ X error and a $50\%$ Z error independently \cite{wu2022erasure,PhysRevLett.102.200501}. The detected leakage information in a dual qubit suggests that there might be propagated errors in the four surrounding primal qubits. The possible error sites and the corresponding propagated errors are illustrated in Fig.\ref{fig2}(b), and we locate the error by adjusting the edge weight, representing the inferred error probabilities, for a matching-based decoder \cite{Higgott2025sparseblossom,wu2023fusionblossomfastmwpm}.\\

\textit{Pure Rydberg decay performance.} Our strategy ensures that the leaked state $\ket{L}$ is not involved in subsequent gates, and exploits final measurement results to locate the propagated error. We first demonstrate the performance of the method for pure Rydberg decay. The results are shown in Fig.\ref{fig3}. First, regarding the threshold, we achieve a high threshold $p_{th} = 3.646\% \approx 3.65\%$, by simulating the logical error rate for a physical error range shown in Fig.\ref{fig3}(a). The threshold for pure Rydberg decay is estimated by fitting the universal scaling ansatz for the critical point of phase transition \cite{WANG200331} with $d = \{7,9,11,13\}$ and an error range from $3.4\%$ to $3.8\%$. For the error distance, we select a physical error range well below the threshold and do linear regression on $\log p_L$ and $\log \frac{p}{p_{th}}$ to derive the slope \cite{supp}.

Indeed, the issue of qubit loss has been addressed previously with MBQC \cite{PhysRevA.90.052316}, but earlier work reported low thresholds and degraded error distances $\lfloor \frac{d+3}{4} \rfloor$ when considering \textit{loss interaction} \cite{PhysRevA.90.052316}. Our strategy makes two major improvements over previous results. First, when the leaked state $\ket{L}$ does not interact during subsequent quantum gates, the leakage error only propagates to correlated Z error. This error propagation pattern is less damaging compared to the previous depolarization-leakage model \cite{PhysRevA.88.042308,suchara2015leakage,brown2020critical,jandura2024surfacecodestabilizermeasurements}. Specifically, it propagates simply as $50\%\ X$ error in the RHG cluster state and does not degrade the error distance. Therefore, our strategy is effective to minimize the damage of Rydberg decay \cite{jandura2024surfacecodestabilizermeasurements}. Second, we have used the \textit{located decoder} to convert propagated errors to located errors, with error distance $d_e \approx d$. These results reveal that the located error with final measurement preserves high threshold and error distance even without exact error location exploited in the erasure conversion protocols \cite{wu2022erasure,sahay2023high}.

\textit{Sub-threshold performance and comparison with biased erasure conversion.} The selection of the QEC strategy strikes a balance between the experimental complexity and error correction performance. We compare our scheme with the best existing erasure conversion protocol \cite{wu2022erasure} for alkaline-earth atoms, biased erasure conversion (BE) \cite{sahay2023high}. The biased erasure conversion uses mid-circuit leakage detection after each two-qubit gate to detect the Rydberg decay error, and then renews each leaked atom with a new atom in the state $\ket{1}$ \cite{sahay2023high}. The resulting channel, after Pauli-twirling, is modeled by $50\%\ Z$ errors. Compared to this scheme, our scheme significantly reduces the requirement by removing the need for mid-circuit leakage detection and site-revolving atom replacement \cite{wu2022erasure,sahay2023high,v7ny-fg31,Chiu_2025,li2025fastcontinuouscoherentatom}. Instead, our scheme only needs a final three-outcome measurement, enabling feasible implementations and broader applicability. 

We then discuss the performance of our strategy and compare it with BE \cite{sahay2023high}. Here, we consider a more general and realistic condition where Pauli errors exist. We assume there are two types of error in the CZ gate \textemdash{} the Rydberg decay, with an error rate $p_e$, and the two-qubit depolarization error, with an error rate $p_d$. The total error rate for a CZ gate is denoted by $p:=p_e+p_d$. We use the two-qubit depolarization error to account for the Rydberg decay back to the qubit-subspace and other noise channels such as atom heating and dephasing \cite{wu2022erasure}. The ratio of Rydberg decay is denoted by $R_e:=p_e/p$ (see \footnote{For $\ce{^{171}_{}Yb}$, an upper limit of $R_e$ is 0.98, estimated by assuming only Rydberg decay exists. For $\ce{^{87}_{}Rb}$, such a limit can be similarly estimated with the data in \cite{evered2023high}, Extended Data Table.1 , Rydberg $T_1=88 \mu s$, $R_e = LG+AL=0.9$, which also validates the range of $R_e$ we select.} for an estimate of the upper limit of $R_e$ for certain alkali and alkaline-earth atoms). Recent experiments have reported $R_e\approx0.55$ \cite{evered2023high} and  $R_e\approx0.56$ \cite{PhysRevX.15.011009}, and the ratio increases when other noises, such as Doppler shifts and laser noise, are further suppressed in future scenarios \footnote{The Rydberg decay branch is fundamentally limited by the finite Rydberg lifetime and cannot be further suppressed. In contrast, other error sources - primarily manifesting as Pauli errors - can be mitigated through atom cooling and pulse shaping.}.

Although our scheme introduces more physically propagated errors than BE, we rigorously show that it achieves better sub-threshold scaling. This is because the additional errors are less harmful \textit{located errors}, leading to a higher proportion of physical located errors. Consequently, our scheme yields comparable, or only slightly degraded, logical error rates at low physical error rates. In Fig.\ref{fig4}(a), we show the comparison in effective error distances, derived from a gate error range $0.1\%$ to $0.5\%$ of interest. The effective error distance for our located decoder (LD) is consistently greater than that of BE, indicating a narrowing gap in logical error rate as the error rate $p$ decreases \cite{supp}. In Fig.\ref{fig4}(b), we show that the gap in logical error rates between these two schemes does not increase apparently when $d$ increases from 5 to 7. Notably, The ratio between two logical error rates $\eta = \frac{p_{L,LD}}{p_{L,BE}}$ is about $1.5$ when $p=0.1\%$ and $d=7$. Furthermore, we prove that when $p$ is small and $d$ is increasing, this ratio is bounded by a constant, by accounting for logical errors introduced by the shortest error chains \cite{Watson_2014,supp}. Although our strategy results in lower thresholds \cite{sahay2023high}, it delivers competitive performance in the sub-threshold regime while also relaxing demanding experimental requirements in the BE protocol. This is particularly relevant given that recent experiments have demonstrated two-qubit gates with error rates well below 1\% \cite{evered2023high,scholl2023erasure,ma2023high,PhysRevX.15.011009} and $R_e$ limited by other noises \cite{ma2023high,PhysRevX.15.011009}. Therefore, our scheme is an alternative and potentially more economical choice to address Rydberg decay, especially in the near term.

\textit{Discussion and conclusion.} To summarize, we have proposed a novel protocol to address Rydberg decay with MBQC without mid-circuit leakage detection and atom replacement. We first mitigate the impact of Rydberg decay efficiently by converting it to a leaked state $\ket{L}$ not involved in subsequent quantum operations and reveal that the error propagation does not degrade the error distance. This is readily achieved by converting BBR errors to loss and isolating RD errors outside the qubit subspace \cite{cong2022hardware,PRXQuantum.5.040343} for different neutral atoms. We second utilize the property of the \textit{located error} to improve the error correction performance, which only exploits a \textit{final three-outcome measurement}. Comparing with the biased erasure conversion \cite{sahay2023high}, our scheme features easier implementation and broader applicability to general neutral atoms, e.g., to the well-established Rb atom platform. Despite achieving lower thresholds, since the additional propagated errors are \textit{located} in our decoding strategy (featured by higher effective distances), our scheme compares favorably with BE at low physical error rates, which is of special interest given recent progress on high-fidelity two-qubit gates \cite{evered2023high,scholl2023erasure,ma2023high,PhysRevX.15.011009}. Therefore, our method serves as an experimentally-friendly alternative to traditional erasure conversion protocols, holding significant implications for fault tolerance on neutral atom platforms \cite{wu2022erasure,sahay2023high}.\\

For handling leakage errors, MBQC is characterized by its ability to refresh qubits via teleportation across different time steps naturally. Consequently, MBQC entails additional qubit overhead and extra two-qubit gates, which lead to slightly degraded logical performance. The RHG cluster takes $1.5\times$ qubit overhead and two-qubit gates compared to the toric code, its circuit counterpart, for the same code distance \cite{Baranes_2026} \footnote{A $d \times d \times d$ RHG cluster state with period boundaries in two directions takes $6d^3$ qubits while a toric code with $d$ rounds of measurement only consumes $(2d^2(data)+2d^2(ancilla))*d = 4d^3$ qubits. A dual qubit in the RHG cluster state connects to 4 primal qubits through 4 CZ gates, and an ancilla qubit connects to 4 data qubits through 4 CZ or CNOT gates. So the number of required two-qubit gates is $12d^3$ for the RHG cluster state and $8d^3$ for the toric code, respectively. We also performed a direct comparison of the overheads for the MBQC scheme versus the (biased) erasure conversion + XZZX code scheme, finding that our scheme incurs an overhead about 2–3 times higher for $R_e=0.7$ \cite{supp}.} . Without incorporating mid-circuit detection and atom replacement — like erasure conversion — the overhead associated with teleportation-based atom renewal is inevitable, but it could be further reduced by exploiting a hybrid framework of MBQC and circuit-based QC \cite{Baranes_2026,Bluvstein_2025}. Another direction to reduce the overhead is to utilize the naturally updated ancilla qubits and dynamical conversion between data qubits and ancilla qubits \cite{zchg-x177,Baranes_2026,eickbusch_demonstration_2025}. Since above methods extend leakage lifetimes and introduce performance trade-offs \footnote{For example, in the RHG cluster state, a qubit is measured after 4 time step, while in a SWAP-Leakage reduction circuit a qubit is measured after 8 time steps \cite{zchg-x177,Baranes_2026,eickbusch_demonstration_2025}}, further investigation is required to identify the exact intervals where they are advantageous compared to MBQC.

Our decoding strategy also effectively handles atom loss in neutral atoms \cite{Baranes_2026,Perrin_2025}. It further extends to other platforms, such as dual-rail qubits \cite{kubica2023erasure,d1v7-nctj,Teoh_2023} and ion-trap qubits \cite{PhysRevA.100.032325,kang2023quantum}, achieving superior logical performance as long as leakage dominates and its propagation does not degrade the distance. For a damaging leakage propagation with degraded distance $d_e=\lfloor \frac{d+3}{4}\rfloor$, locating propagated errors via final measurements ensures $d_e = \frac{d+1}{2}$, which inspires hardware-efficient protocols for handling the leakage \cite{zchg-x177}. Regarding future decoder optimization, we expect that decoding accuracy would benefit from a more accurate inference of error probabilities based on detected leakage. This could potentially be realized via BP-enhanced decoders \cite{PhysRevX.13.031007} or machine learning decoders \cite{Bluvstein_2025}. Furthermore, we anticipate that our method can be extended to decoding logical operations, thereby facilitating broader applications \cite{Baranes_2026,PhysRevLett.133.240602,cain2025fastcorrelateddecodingtransversal}.\\

\textit{Code availability.} The code for simulation is available in \cite{code}.\\

\textit{Acknowledgments}---We acknowledge helpful discussions with Yuchen Zhang, Haowen Cheng, Zhaoqiu Zengxu, Pai Peng, Lin Li, Hugo Perrin, Vikas Buchemmavari, Gefen Baranes and Pengyu Liu. This work was supported by the HFNL Self-Deployed Project (Nos. ZB2024010101, ZB2024010102, ZB2024010201, and ZB2024010501), the Quantum Science and Technology-National Science and Technology Major Project (No. 2021ZD0301405), the National Key R\&D Program of China (No. 2022ZD0160101), the Shanghai Municipal Science and Technology Commission (No. 24DP2600300), the National Natural Science Foundation of China (No. 12322415), and the New Cornerstone Science Foundation.

\bibliographystyle{apsrev4-2}
\bibliography{ref}

@article{Baranes_2026,
   title={Leveraging Qubit Loss Detection in Fault-Tolerant Quantum Algorithms},
   volume={16},
   ISSN={2160-3308},
   url={http://dx.doi.org/10.1103/ycwc-3myc},
   DOI={10.1103/ycwc-3myc},
   number={1},
   journal={Physical Review X},
   publisher={American Physical Society (APS)},
   author={Baranes, Gefen and Cain, Madelyn and Ataides, J. Pablo Bonilla and Bluvstein, Dolev and Sinclair, Josiah and Vuletić, Vladan and Zhou, Hengyun and Lukin, Mikhail D.},
   year={2026},
   month=jan }

@misc{wu2023fusionblossomfastmwpm,
      title={Fusion Blossom: Fast MWPM Decoders for QEC}, 
      author={Yue Wu and Lin Zhong},
      year={2023},
      eprint={2305.08307},
      archivePrefix={arXiv},
      primaryClass={quant-ph},
      url={https://arxiv.org/abs/2305.08307}, 
}

@misc{jandura2024surfacecodestabilizermeasurements,
      title={Surface Code Stabilizer Measurements for Rydberg Atoms}, 
      author={Sven Jandura and Guido Pupillo},
      year={2024},
      eprint={2405.16621},
      archivePrefix={arXiv},
      primaryClass={quant-ph},
      url={https://arxiv.org/abs/2405.16621}, 
}

@misc{omanakuttan2024coherencepreservingleakagedetection,
      title={Coherence Preserving Leakage Detection and Cooling in Alkaline Earth Atoms}, 
      author={Sivaprasad Omanakuttan and Vikas Buchemmavari and Michael J. Martin and Ivan H Deutsch},
      year={2024},
      eprint={2410.23430},
      archivePrefix={arXiv},
      primaryClass={quant-ph},
      url={https://arxiv.org/abs/2410.23430}, 
}

@article{evered2023high,
  title={High-fidelity parallel entangling gates on a neutral-atom quantum computer},
  author={Evered, Simon J and Bluvstein, Dolev and Kalinowski, Marcin and Ebadi, Sepehr and Manovitz, Tom and Zhou, Hengyun and Li, Sophie H and Geim, Alexandra A and Wang, Tout T and Maskara, Nishad and others},
  journal={Nature},
  volume={622},
  number={7982},
  pages={268--272},
  year={2023},
  publisher={Nature Publishing Group UK London}
}

@article{scholl2023erasure,
  title={Erasure conversion in a high-fidelity Rydberg quantum simulator},
  author={Scholl, Pascal and Shaw, Adam L and Tsai, Richard Bing-Shiun and Finkelstein, Ran and Choi, Joonhee and Endres, Manuel},
  journal={Nature},
  volume={622},
  number={7982},
  pages={273--278},
  year={2023},
  publisher={Nature Publishing Group UK London}
}

@article{Watson_2014,
   title={Logical error rate scaling of the toric code},
   volume={16},
   ISSN={1367-2630},
   url={http://dx.doi.org/10.1088/1367-2630/16/9/093045},
   DOI={10.1088/1367-2630/16/9/093045},
   number={9},
   journal={New Journal of Physics},
   publisher={IOP Publishing},
   author={Watson, Fern H E and Barrett, Sean D},
   year={2014},
   month=sep, pages={093045} }

@article{Ebadi_2021,
   title={Quantum phases of matter on a 256-atom programmable quantum simulator},
   volume={595},
   ISSN={1476-4687},
   url={http://dx.doi.org/10.1038/s41586-021-03582-4},
   DOI={10.1038/s41586-021-03582-4},
   number={7866},
   journal={Nature},
   publisher={Springer Science and Business Media LLC},
   author={Ebadi, Sepehr and Wang, Tout T. and Levine, Harry and Keesling, Alexander and Semeghini, Giulia and Omran, Ahmed and Bluvstein, Dolev and Samajdar, Rhine and Pichler, Hannes and Ho, Wen Wei and Choi, Soonwon and Sachdev, Subir and Greiner, Markus and Vuletić, Vladan and Lukin, Mikhail D.},
   year={2021},
   month=jul, pages={227–232} }

@article{bluvstein2024logical,
  title={Logical quantum processor based on reconfigurable atom arrays},
  author={Bluvstein, Dolev and Evered, Simon J and Geim, Alexandra A and Li, Sophie H and Zhou, Hengyun and Manovitz, Tom and Ebadi, Sepehr and Cain, Madelyn and Kalinowski, Marcin and Hangleiter, Dominik and others},
  journal={Nature},
  volume={626},
  number={7997},
  pages={58--65},
  year={2024},
  publisher={Nature Publishing Group UK London}
}

@book{Nielsen_Chuang_2010, 
      place={Cambridge}, 
      title={Quantum Computation and Quantum Information: 10th Anniversary Edition}, publisher={Cambridge University Press}, 
      author={Nielsen, Michael A. and Chuang, Isaac L.}, 
      year={2010}}

@misc{fowler2009topologicalclusterstatequantum,
      title={Topological cluster state quantum computing}, 
      author={Austin G. Fowler and Kovid Goyal},
      year={2009},
      eprint={0805.3202},
      archivePrefix={arXiv},
      primaryClass={quant-ph},
      url={https://arxiv.org/abs/0805.3202}, 
}

@article{ma2023high,
  title={High-fidelity gates and mid-circuit erasure conversion in an atomic qubit},
  author={Ma, Shuo and Liu, Genyue and Peng, Pai and Zhang, Bichen and Jandura, Sven and Claes, Jahan and Burgers, Alex P and Pupillo, Guido and Puri, Shruti and Thompson, Jeff D},
  journal={Nature},
  volume={622},
  number={7982},
  pages={279--284},
  year={2023},
  publisher={Nature Publishing Group UK London}
}

@article{PhysRevA.90.052316,
  title = {Upper bound for loss in practical topological-cluster-state quantum computing},
  author = {Whiteside, Adam C. and Fowler, Austin G.},
  journal = {Phys. Rev. A},
  volume = {90},
  issue = {5},
  pages = {052316},
  numpages = {11},
  year = {2014},
  month = {Nov},
  publisher = {American Physical Society},
  doi = {10.1103/PhysRevA.90.052316},
  url = {https://link.aps.org/doi/10.1103/PhysRevA.90.052316}
}

@article{jayashankar2022achieving,
  title={Achieving fault tolerance against amplitude-damping noise},
  author={Jayashankar, Akshaya and Long, My Duy Hoang and Ng, Hui Khoon and Mandayam, Prabha},
  journal={Physical Review Research},
  volume={4},
  number={2},
  pages={023034},
  year={2022},
  publisher={APS}
}

@article{Wallman_2016,
   title={Noise tailoring for scalable quantum computation via randomized compiling},
   volume={94},
   ISSN={2469-9934},
   url={http://dx.doi.org/10.1103/PhysRevA.94.052325},
   DOI={10.1103/physreva.94.052325},
   number={5},
   journal={Physical Review A},
   publisher={American Physical Society (APS)},
   author={Wallman, Joel J. and Emerson, Joseph},
   year={2016},
   month=nov }

@article{d1v7-nctj,
  title = {Surface Code with Imperfect Erasure Checks},
  author = {Chang, Kathleen and Singh, Shraddha and Claes, Jahan and Sahay, Kaavya and Teoh, James and Puri, Shruti},
  journal = {PRX Quantum},
  volume = {6},
  issue = {4},
  pages = {040355},
  numpages = {21},
  year = {2025},
  month = {Dec},
  publisher = {American Physical Society},
  doi = {10.1103/d1v7-nctj},
  url = {https://link.aps.org/doi/10.1103/d1v7-nctj}
}

@article{PhysRevX.13.031007,
  title = {Improved Decoding of Circuit Noise and Fragile Boundaries of Tailored Surface Codes},
  author = {Higgott, Oscar and Bohdanowicz, Thomas C. and Kubica, Aleksander and Flammia, Steven T. and Campbell, Earl T.},
  journal = {Phys. Rev. X},
  volume = {13},
  issue = {3},
  pages = {031007},
  numpages = {20},
  year = {2023},
  month = {Jul},
  publisher = {American Physical Society},
  doi = {10.1103/PhysRevX.13.031007},
  url = {https://link.aps.org/doi/10.1103/PhysRevX.13.031007}
}

@article{PhysRevApplied.19.034089,
  title = {Complete Unitary Qutrit Control in Ultracold Atoms},
  author = {Lindon, Joseph and Tashchilina, Arina and Cooke, Logan W. and LeBlanc, Lindsay J.},
  journal = {Phys. Rev. Appl.},
  volume = {19},
  issue = {3},
  pages = {034089},
  numpages = {11},
  year = {2023},
  month = {Mar},
  publisher = {American Physical Society},
  doi = {10.1103/PhysRevApplied.19.034089},
  url = {https://link.aps.org/doi/10.1103/PhysRevApplied.19.034089}
}

@article{kubica2023erasure,
  title={Erasure qubits: Overcoming the T 1 limit in superconducting circuits},
  author={Kubica, Aleksander and Haim, Arbel and Vaknin, Yotam and Levine, Harry and Brand{\~a}o, Fernando and Retzker, Alex},
  journal={Physical Review X},
  volume={13},
  number={4},
  pages={041022},
  year={2023},
  publisher={APS}
}

@article{PhysRevA.100.032325,
  title = {Leakage mitigation for quantum error correction using a mixed qubit scheme},
  author = {Brown, Natalie C. and Brown, Kenneth R.},
  journal = {Phys. Rev. A},
  volume = {100},
  issue = {3},
  pages = {032325},
  numpages = {9},
  year = {2019},
  month = {Sep},
  publisher = {American Physical Society},
  doi = {10.1103/PhysRevA.100.032325},
  url = {https://link.aps.org/doi/10.1103/PhysRevA.100.032325}
}

@article{Bravyi_2018,
   title={Correcting coherent errors with surface codes},
   volume={4},
   ISSN={2056-6387},
   url={http://dx.doi.org/10.1038/s41534-018-0106-y},
   DOI={10.1038/s41534-018-0106-y},
   number={1},
   journal={npj Quantum Information},
   publisher={Springer Science and Business Media LLC},
   author={Bravyi, Sergey and Englbrecht, Matthias and König, Robert and Peard, Nolan},
   year={2018},
   month=oct }

@article{PhysRevX.13.041051,
  title = {Midcircuit Measurements on a Single-Species Neutral Alkali Atom Quantum Processor},
  author = {Graham, T. M. and Phuttitarn, L. and Chinnarasu, R. and Song, Y. and Poole, C. and Jooya, K. and Scott, J. and Scott, A. and Eichler, P. and Saffman, M.},
  journal = {Phys. Rev. X},
  volume = {13},
  issue = {4},
  pages = {041051},
  numpages = {22},
  year = {2023},
  month = {Dec},
  publisher = {American Physical Society},
  doi = {10.1103/PhysRevX.13.041051},
  url = {https://link.aps.org/doi/10.1103/PhysRevX.13.041051}
}

@article{PhysRevLett.86.5188,
  title = {A One-Way Quantum Computer},
  author = {Raussendorf, Robert and Briegel, Hans J.},
  journal = {Phys. Rev. Lett.},
  volume = {86},
  issue = {22},
  pages = {5188--5191},
  numpages = {0},
  year = {2001},
  month = {May},
  publisher = {American Physical Society},
  doi = {10.1103/PhysRevLett.86.5188},
  url = {https://link.aps.org/doi/10.1103/PhysRevLett.86.5188}
}

@article{briegel2009measurement,
  title={Measurement-based quantum computation},
  author={Briegel, Hans J and Browne, David E and D{\"u}r, Wolfgang and Raussendorf, Robert and Van den Nest, Maarten},
  journal={Nature Physics},
  volume={5},
  number={1},
  pages={19--26},
  year={2009},
  publisher={Nature Publishing Group UK London}
}

@article{PhysRevA.68.022312,
  title = {Measurement-based quantum computation on cluster states},
  author = {Raussendorf, Robert and Browne, Daniel E. and Briegel, Hans J.},
  journal = {Phys. Rev. A},
  volume = {68},
  issue = {2},
  pages = {022312},
  numpages = {32},
  year = {2003},
  month = {Aug},
  publisher = {American Physical Society},
  doi = {10.1103/PhysRevA.68.022312},
  url = {https://link.aps.org/doi/10.1103/PhysRevA.68.022312}
}

@article{zchg-x177,
  title = {Locating Rydberg decay error in the SWAP-Leakage Reduction Circuit protocol},
  author = {Yu, Cheng-Cheng and Deng, Yu-Hao and Lu, Chao-Yang and Chen, Ming-Cheng and Pan, Jian-Wei},
  journal = {Phys. Rev. A},
  pages = {--},
  year = {2025},
  month = {Sep},
  publisher = {American Physical Society},
  doi = {10.1103/zchg-x177},
  url = {https://link.aps.org/doi/10.1103/zchg-x177}
}

@article{raussendorf2006fault,
  title={A fault-tolerant one-way quantum computer},
  author={Raussendorf, Robert and Harrington, Jim and Goyal, Kovid},
  journal={Annals of physics},
  volume={321},
  number={9},
  pages={2242--2270},
  year={2006},
  publisher={Elsevier}
}

@article{raussendorf2007topological,
  title={Topological fault-tolerance in cluster state quantum computation},
  author={Raussendorf, Robert and Harrington, Jim and Goyal, Kovid},
  journal={New Journal of Physics},
  volume={9},
  number={6},
  pages={199},
  year={2007},
  publisher={IOP Publishing}
}

@article{PhysRevLett.98.190504,
  title = {Fault-Tolerant Quantum Computation with High Threshold in Two Dimensions},
  author = {Raussendorf, Robert and Harrington, Jim},
  journal = {Phys. Rev. Lett.},
  volume = {98},
  issue = {19},
  pages = {190504},
  numpages = {4},
  year = {2007},
  month = {May},
  publisher = {American Physical Society},
  doi = {10.1103/PhysRevLett.98.190504},
  url = {https://link.aps.org/doi/10.1103/PhysRevLett.98.190504}
}

@article{PhysRevLett.117.070501,
  title = {Foliated Quantum Error-Correcting Codes},
  author = {Bolt, A. and Duclos-Cianci, G. and Poulin, D. and Stace, T. M.},
  journal = {Phys. Rev. Lett.},
  volume = {117},
  issue = {7},
  pages = {070501},
  numpages = {6},
  year = {2016},
  month = {Aug},
  publisher = {American Physical Society},
  doi = {10.1103/PhysRevLett.117.070501},
  url = {https://link.aps.org/doi/10.1103/PhysRevLett.117.070501}
}

@article{PhysRevResearch.2.033305,
  title = {Universal fault-tolerant measurement-based quantum computation},
  author = {Brown, Benjamin J. and Roberts, Sam},
  journal = {Phys. Rev. Res.},
  volume = {2},
  issue = {3},
  pages = {033305},
  numpages = {27},
  year = {2020},
  month = {Aug},
  publisher = {American Physical Society},
  doi = {10.1103/PhysRevResearch.2.033305},
  url = {https://link.aps.org/doi/10.1103/PhysRevResearch.2.033305}
}

@article{sahay2023high,
  title={High-threshold codes for neutral-atom qubits with biased erasure errors},
  author={Sahay, Kaavya and Jin, Junlan and Claes, Jahan and Thompson, Jeff D and Puri, Shruti},
  journal={Physical Review X},
  volume={13},
  number={4},
  pages={041013},
  year={2023},
  publisher={APS}
}

@article{cong2022hardware,
  title={Hardware-efficient, fault-tolerant quantum computation with Rydberg atoms},
  author={Cong, Iris and Levine, Harry and Keesling, Alexander and Bluvstein, Dolev and Wang, Sheng-Tao and Lukin, Mikhail D},
  journal={Physical Review X},
  volume={12},
  number={2},
  pages={021049},
  year={2022},
  publisher={APS}
}

@article{PhysRevA.88.042308,
  title = {Coping with qubit leakage in topological codes},
  author = {Fowler, Austin G.},
  journal = {Phys. Rev. A},
  volume = {88},
  issue = {4},
  pages = {042308},
  numpages = {5},
  year = {2013},
  month = {Oct},
  publisher = {American Physical Society},
  doi = {10.1103/PhysRevA.88.042308},
  url = {https://link.aps.org/doi/10.1103/PhysRevA.88.042308}
}

@article{WANG200331,
title = {Confinement-Higgs transition in a disordered gauge theory and the accuracy threshold for quantum memory},
journal = {Annals of Physics},
volume = {303},
number = {1},
pages = {31-58},
year = {2003},
issn = {0003-4916},
doi = {https://doi.org/10.1016/S0003-4916(02)00019-2},
url = {https://www.sciencedirect.com/science/article/pii/S0003491602000192},
author = {Chenyang Wang and Jim Harrington and John Preskill},
}

@inproceedings{suchara2015leakage,
  title={Leakage suppression in the toric code},
  author={Suchara, Martin and Cross, Andrew W and Gambetta, Jay M},
  booktitle={2015 IEEE International Symposium on Information Theory (ISIT)},
  pages={1119--1123},
  year={2015},
  organization={IEEE}
}

@inproceedings{brown2020critical,
  title={Critical faults of leakage errors on the surface code},
  author={Brown, Natalie C and Cross, Andrew and Brown, Kenneth R},
  booktitle={2020 IEEE International Conference on Quantum Computing and Engineering (QCE)},
  pages={286--294},
  year={2020},
  organization={IEEE}
}

@article{bluvstein2022quantum,
  title={A quantum processor based on coherent transport of entangled atom arrays},
  author={Bluvstein, Dolev and Levine, Harry and Semeghini, Giulia and Wang, Tout T and Ebadi, Sepehr and Kalinowski, Marcin and Keesling, Alexander and Maskara, Nishad and Pichler, Hannes and Greiner, Markus and others},
  journal={Nature},
  volume={604},
  number={7906},
  pages={451--456},
  year={2022},
  publisher={Nature Publishing Group UK London}
}

@article{Perrin_2025,
   title={Quantum Error Correction resilient against Atom Loss},
   volume={9},
   ISSN={2521-327X},
   url={http://dx.doi.org/10.22331/q-2025-10-13-1884},
   DOI={10.22331/q-2025-10-13-1884},
   journal={Quantum},
   publisher={Verein zur Forderung des Open Access Publizierens in den Quantenwissenschaften},
   author={Perrin, Hugo and Jandura, Sven and Pupillo, Guido},
   year={2025},
   month=oct, pages={1884} }

@article{PhysRevX.15.011009,
  title = {Spectroscopy and Modeling of $^{171}\mathrm{Yb}$ Rydberg States for High-Fidelity Two-Qubit Gates},
  author = {Peper, Michael and Li, Yiyi and Knapp, Daniel Y. and Bileska, Mila and Ma, Shuo and Liu, Genyue and Peng, Pai and Zhang, Bichen and Horvath, Sebastian P. and Burgers, Alex P. and Thompson, Jeff D.},
  journal = {Phys. Rev. X},
  volume = {15},
  issue = {1},
  pages = {011009},
  numpages = {30},
  year = {2025},
  month = {Jan},
  publisher = {American Physical Society},
  doi = {10.1103/PhysRevX.15.011009},
  url = {https://link.aps.org/doi/10.1103/PhysRevX.15.011009}
}

@article{PhysRevLett.102.200501,
  title = {Thresholds for Topological Codes in the Presence of Loss},
  author = {Stace, Thomas M. and Barrett, Sean D. and Doherty, Andrew C.},
  journal = {Phys. Rev. Lett.},
  volume = {102},
  issue = {20},
  pages = {200501},
  numpages = {4},
  year = {2009},
  month = {May},
  publisher = {American Physical Society},
  doi = {10.1103/PhysRevLett.102.200501},
  url = {https://link.aps.org/doi/10.1103/PhysRevLett.102.200501}
}

@article{Teoh_2023,
   title={Dual-rail encoding with superconducting cavities},
   volume={120},
   ISSN={1091-6490},
   url={http://dx.doi.org/10.1073/pnas.2221736120},
   DOI={10.1073/pnas.2221736120},
   number={41},
   journal={Proceedings of the National Academy of Sciences},
   publisher={Proceedings of the National Academy of Sciences},
   author={Teoh, James D. and Winkel, Patrick and Babla, Harshvardhan K. and Chapman, Benjamin J. and Claes, Jahan and de Graaf, Stijn J. and Garmon, John W. O. and Kalfus, William D. and Lu, Yao and Maiti, Aniket and Sahay, Kaavya and Thakur, Neel and Tsunoda, Takahiro and Xue, Sophia H. and Frunzio, Luigi and Girvin, Steven M. and Puri, Shruti and Schoelkopf, Robert J.},
   year={2023},
   month=oct }

@article{PhysRevLett.105.200502,
  title = {Fault Tolerant Quantum Computation with Very High Threshold for Loss Errors},
  author = {Barrett, Sean D. and Stace, Thomas M.},
  journal = {Phys. Rev. Lett.},
  volume = {105},
  issue = {20},
  pages = {200502},
  numpages = {4},
  year = {2010},
  month = {Nov},
  publisher = {American Physical Society},
  doi = {10.1103/PhysRevLett.105.200502},
  url = {https://link.aps.org/doi/10.1103/PhysRevLett.105.200502}
}

@article{wu2022erasure,
  title={Erasure conversion for fault-tolerant quantum computing in alkaline earth Rydberg atom arrays},
  author={Wu, Yue and Kolkowitz, Shimon and Puri, Shruti and Thompson, Jeff D},
  journal={Nature communications},
  volume={13},
  number={1},
  pages={4657},
  year={2022},
  publisher={Nature Publishing Group UK London}
}

@article{kang2023quantum,
  title={Quantum error correction with metastable states of trapped ions using erasure conversion},
  author={Kang, Mingyu and Campbell, Wesley C and Brown, Kenneth R},
  journal={PRX Quantum},
  volume={4},
  number={2},
  pages={020358},
  year={2023},
  publisher={APS}
}

@article{Gidney_2021,
   title={Stim: a fast stabilizer circuit simulator},
   volume={5},
   ISSN={2521-327X},
   url={http://dx.doi.org/10.22331/q-2021-07-06-497},
   DOI={10.22331/q-2021-07-06-497},
   journal={Quantum},
   publisher={Verein zur Forderung des Open Access Publizierens in den Quantenwissenschaften},
   author={Gidney, Craig},
   year={2021},
   month=jul, pages={497} }

@article{PhysRevResearch.7.013249,
  title = {Fault-tolerant quantum architectures based on erasure qubits},
  author = {Gu, Shouzhen and Retzker, Alex and Kubica, Aleksander},
  journal = {Phys. Rev. Res.},
  volume = {7},
  issue = {1},
  pages = {013249},
  numpages = {17},
  year = {2025},
  month = {Mar},
  publisher = {American Physical Society},
  doi = {10.1103/PhysRevResearch.7.013249},
  url = {https://link.aps.org/doi/10.1103/PhysRevResearch.7.013249}
}

@InProceedings{ seabold-proc-scipy-2010,
  author    = { {S}kipper {S}eabold and {J}osef {P}erktold },
  title     = { {S}tatsmodels: {E}conometric and {S}tatistical {M}odeling with {P}ython },
  booktitle = { {P}roceedings of the 9th {P}ython in {S}cience {C}onference },
  pages     = { 92 - 96 },
  year      = { 2010 },
  editor    = { {S}t\'efan van der {W}alt and {J}arrod {M}illman },
  doi       = { 10.25080/Majora-92bf1922-011 }
}

@article{aliferis2007fault,
  title={Fault-tolerant quantum computation for local leakage faults},
  author={Aliferis, Panos and Terhal, Barbara M},
  journal={Quantum Information and Computation},
  volume={7},
  number={1-2},
  pages={139--156},
  year={2007},
  publisher={Rinton Press Inc.}
}

@misc{cain2025fastcorrelateddecodingtransversal,
      title={Fast correlated decoding of transversal logical algorithms}, 
      author={Madelyn Cain and Dolev Bluvstein and Chen Zhao and Shouzhen Gu and Nishad Maskara and Marcin Kalinowski and Alexandra A. Geim and Aleksander Kubica and Mikhail D. Lukin and Hengyun Zhou},
      year={2025},
      eprint={2505.13587},
      archivePrefix={arXiv},
      primaryClass={quant-ph},
      url={https://arxiv.org/abs/2505.13587}, 
}

@article{Zhou_2025,
   title={Low-overhead transversal fault tolerance for universal quantum computation},
   volume={646},
   ISSN={1476-4687},
   url={http://dx.doi.org/10.1038/s41586-025-09543-5},
   DOI={10.1038/s41586-025-09543-5},
   number={8084},
   journal={Nature},
   publisher={Springer Science and Business Media LLC},
   author={Zhou, Hengyun and Zhao, Chen and Cain, Madelyn and Bluvstein, Dolev and Maskara, Nishad and Duckering, Casey and Hu, Hong-Ye and Wang, Sheng-Tao and Kubica, Aleksander and Lukin, Mikhail D.},
   year={2025},
   month=sep, pages={303–308} }

@article{Tournaire_2026,
   title={A 3D lattice defect and efficient computations in topological MBQC},
   volume={10},
   ISSN={2521-327X},
   url={http://dx.doi.org/10.22331/q-2026-02-06-1997},
   DOI={10.22331/q-2026-02-06-1997},
   journal={Quantum},
   publisher={Verein zur Forderung des Open Access Publizierens in den Quantenwissenschaften},
   author={Tournaire, Gabrielle and Schwiering, Marvin and Raussendorf, Robert and Bachmann, Sven},
   year={2026},
   month=feb, pages={1997} }

@article{Dennis_2002,
   title={Topological quantum memory},
   volume={43},
   ISSN={1089-7658},
   url={http://dx.doi.org/10.1063/1.1499754},
   DOI={10.1063/1.1499754},
   number={9},
   journal={Journal of Mathematical Physics},
   publisher={AIP Publishing},
   author={Dennis, Eric and Kitaev, Alexei and Landahl, Andrew and Preskill, John},
   year={2002},
   month=sep, pages={4452–4505} }

@Misc{code,
howpublished = {\url{https://github.com/yuchengcheng720/Located-decoder-for-Rydberg-decay}},
title = {Available code for this work}
}

@misc{li2025fastcontinuouscoherentatom,
      title={Fast, continuous and coherent atom replacement in a neutral atom qubit array}, 
      author={Yiyi Li and Yicheng Bao and Michael Peper and Chenyuan Li and Jeff D. Thompson},
      year={2025},
      eprint={2506.15633},
      archivePrefix={arXiv},
      primaryClass={quant-ph},
      url={https://arxiv.org/abs/2506.15633}, 
}

@article{Chiu_2025,
   title={Continuous operation of a coherent 3,000-qubit system},
   ISSN={1476-4687},
   url={http://dx.doi.org/10.1038/s41586-025-09596-6},
   DOI={10.1038/s41586-025-09596-6},
   journal={Nature},
   publisher={Springer Science and Business Media LLC},
   author={Chiu, Neng-Chun and Trapp, Elias C. and Guo, Jinen and Abobeih, Mohamed H. and Stewart, Luke M. and Hollerith, Simon and Stroganov, Pavel L. and Kalinowski, Marcin and Geim, Alexandra A. and Evered, Simon J. and Li, Sophie H. and Lyu, Xingjian and Peters, Lisa M. and Bluvstein, Dolev and Wang, Tout T. and Greiner, Markus and Vuletić, Vladan and Lukin, Mikhail D.},
   year={2025},
   month=sep }

@article{v7ny-fg31,
  title = {Repeated Ancilla Reuse for Logical Computation on a Neutral Atom Quantum Computer},
  author = {Muniz, J. A. and Crow, D. and Kim, H. and Kindem, J. M. and Cairncross, W. B. and Ryou, A. and Bohdanowicz, T. C. and Chen, C.-A. and Ji, Y. and Jones, A. M. W. and Megidish, E. and Nishiguchi, C. and Urbanek, M. and Wadleigh, L. and Wilkason, T. and Aasen, D. and Barnes, K. and Bello-Rivas, J. M. and Bloomfield, I. and Booth, G. and Brown, A. and Brown, M. O. and Cassella, K. and Cowan, G. and Epstein, J. and Feldkamp, M. and Griger, C. and Hassan, Y. and Heinz, A. and Halperin, E. and Hofler, T. and Hummel, F. and Jaffe, M. and Kapit, E. and Kotru, K. and Lauigan, J. and Marjanovic, J. and Meredith, M. and McDonald, M. and Morshead, R. and Narayanaswami, S. and Pawlak, K. A. and Pudenz, K. L. and P\'erez, D. Rodr\'{\i}guez and Sabharwal, P. and Simon, J. and Smull, A. and Sorensen, M. and Stack, D. T. and Stone, M. and Taneja, L. and van de Veerdonk, R. J. M. and Vendeiro, Z. and Weverka, R. T. and White, K. and Wu, T.-Y. and Xie, X. and Zalys-Geller, E. and Zhang, X. and King, J. and Bloom, B. J. and Norcia, M. A.},
  journal = {Phys. Rev. X},
  volume = {15},
  issue = {4},
  pages = {041040},
  numpages = {16},
  year = {2025},
  month = {Dec},
  publisher = {American Physical Society},
  doi = {10.1103/v7ny-fg31},
  url = {https://link.aps.org/doi/10.1103/v7ny-fg31}
}

@article{PhysRevLett.133.240602,
  title = {Correlated Decoding of Logical Algorithms with Transversal Gates},
  author = {Cain, Madelyn and Zhao, Chen and Zhou, Hengyun and Meister, Nadine and Ataides, J. Pablo Bonilla and Jaffe, Arthur and Bluvstein, Dolev and Lukin, Mikhail D.},
  journal = {Phys. Rev. Lett.},
  volume = {133},
  issue = {24},
  pages = {240602},
  numpages = {7},
  year = {2024},
  month = {Dec},
  publisher = {American Physical Society},
  doi = {10.1103/PhysRevLett.133.240602},
  url = {https://link.aps.org/doi/10.1103/PhysRevLett.133.240602}
}

@article{Higgott2025sparseblossom,
  doi = {10.22331/q-2025-01-20-1600},
  url = {https://doi.org/10.22331/q-2025-01-20-1600},
  title = {Sparse {B}lossom: correcting a million errors per core second with minimum-weight matching},
  author = {Higgott, Oscar and Gidney, Craig},
  journal = {{Quantum}},
  issn = {2521-327X},
  publisher = {{Verein zur F{\"{o}}rderung des Open Access Publizierens in den Quantenwissenschaften}},
  volume = {9},
  pages = {1600},
  month = jan,
  year = {2025}
}

@misc{Gidney2024PyMatchingEdges,
  author       = {Gidney, Craig},
  title        = {Answer to ``Remove edges from {PyMatching} matching graph / {Copy} {Matching} object''},
  howpublished = {Quantum Computing Stack Exchange},
  year         = {2024},
  url          = {https://quantumcomputing.stackexchange.com/questions/38217/remove-edges-from-pymatching-matching-graph-copy-matching-object},
  note         = {Accessed: 2025-12-24}
}

@article{eickbusch_demonstration_2025,
	title = {Demonstration of dynamic surface codes},
	volume = {21},
	doi = {10.1038/s41567-025-03070-w},
	number = {12},
	journal = {Nature Physics},
	author = {Eickbusch, Alec and McEwen, Matt and others},
	year = {2025},
	pages = {1994--2001},
}

@article{Bluvstein_2025,
   title={A fault-tolerant neutral-atom architecture for universal quantum computation},
   volume={649},
   ISSN={1476-4687},
   url={http://dx.doi.org/10.1038/s41586-025-09848-5},
   DOI={10.1038/s41586-025-09848-5},
   number={8095},
   journal={Nature},
   publisher={Springer Science and Business Media LLC},
   author={Bluvstein, Dolev and Geim, Alexandra A. and Li, Sophie H. and Evered, Simon J. and Bonilla Ataides, J. Pablo and Baranes, Gefen and Gu, Andi and Manovitz, Tom and Xu, Muqing and Kalinowski, Marcin and Majidy, Shayan and Kokail, Christian and Maskara, Nishad and Trapp, Elias C. and Stewart, Luke M. and Hollerith, Simon and Zhou, Hengyun and Gullans, Michael J. and Yelin, Susanne F. and Greiner, Markus and Vuletić, Vladan and Cain, Madelyn and Lukin, Mikhail D.},
   year={2025},
   month=nov, pages={39–46} }

@misc{wu2025minimumweightparityfactordecoder,
      title={Minimum-Weight Parity Factor Decoder for Quantum Error Correction}, 
      author={Yue Wu and Binghong Li and Kathleen Chang and Shruti Puri and Lin Zhong},
      year={2025},
      eprint={2508.04969},
      archivePrefix={arXiv},
      primaryClass={quant-ph},
      url={https://arxiv.org/abs/2508.04969}, 
}

@article{PRXQuantum.5.040343,
  title = {Circuit-Based Leakage-to-Erasure Conversion in a Neutral-Atom Quantum Processor},
  author = {Chow, Matthew N. H. and Buchemmavari, Vikas and Omanakuttan, Sivaprasad and Little, Bethany J. and Pandey, Saurabh and Deutsch, Ivan H. and Jau, Yuan-Yu},
  journal = {PRX Quantum},
  volume = {5},
  issue = {4},
  pages = {040343},
  numpages = {14},
  year = {2024},
  month = {Dec},
  publisher = {American Physical Society},
  doi = {10.1103/PRXQuantum.5.040343},
  url = {https://link.aps.org/doi/10.1103/PRXQuantum.5.040343}
}

@article{985g-58gd,
  title = {Optimizing Quantum Error-Correction Protocols with Erasure Qubits},
  author = {Gu, Shouzhen and Vaknin, Yotam and Retzker, Alex and Kubica, Aleksander},
  journal = {PRX Quantum},
  volume = {6},
  issue = {4},
  pages = {040354},
  numpages = {21},
  year = {2025},
  month = {Dec},
  publisher = {American Physical Society},
  doi = {10.1103/985g-58gd},
  url = {https://link.aps.org/doi/10.1103/985g-58gd}
}

@misc{supp,
  title        = {See {S}upplemental {M}aterial for experimental justification and derivation of the error propagation property, {RHG} cluster state and {CZ} gate sequence, numerical simulation details and extended discussion, logical error rates comparison at low physical error rates, and comparison with (biased) erasure conversion in {XZZX} codes.},
}

\end{document}

\maketitle
\begin{center}
\textbf{\large --- Supplemental Material ---\\[0.5em]Taming Rydberg Decay with Measurement-based Quantum Computation}\\[1em]
    Cheng-Cheng Yu$^{1,2,3}$, Zi-Han Chen$^{1,2,3}$, Yu-Hao Deng$^{1,2,3}$, Chao-Yang Lu$^{1,2,3,4}$, Ming-Cheng Chen$^{1,2,3}$, and Jian-Wei Pan$^{1,2,3}$

	{\centering \emph{$^1$Hefei National Research Center for Physical Sciences at the Microscale and School of Physical Sciences, University of Science and Technology of China, Hefei 230026, China}}
	
	{\centering \emph{$^2$Shanghai Research Center for Quantum Science and CAS Center for Excellence in Quantum Information and Quantum Physics, University of Science and Technology of China, Shanghai 201315, China}}
	
	{\centering \emph{$^3$Hefei National Laboratory, University of Science and Technology of China, Hefei 230088, China}}
	
	{\centering \emph{$^4$New Cornerstone Science Laboratory, Hefei, 230026, China}}

	\thispagestyle{titlepage}
\end{center}
\renewcommand{\thesection}{S.\Roman{section}}
	
The contents of the supplementary material are structured as follows: In \ref{appA}, we discuss the error model of the Rydberg decay error, including the hardware justification, how to use Pauli-twirling to remove potentially harmful coherent errors, the resulting error model, and the property of error propagation. In \ref{appB}, we elaborate on the structure of the RHG cluster state and discuss the propagation of the Rydberg decay error. In \ref{appC}, we have provided an extended discussion and technical details regarding the numerical simulations. In \ref{appD}, we provide an extended discussion about comparison in terms of the logical error rate, in the small error rate limit with increasing distance. In \ref{appE}, we compared the performance of our scheme with circuit-based erasure conversion (on the XZZX code) and estimated the resource overhead. In \ref{appF}, we benchmarked the runtime of the error correction simulation algorithm and discussed directions for future improvement. In \ref{appG}, we compare our work with several relevant recent works.

\section{Rydberg Decay Error Model}\label{appA}
\subsection{Hardware Justification for the Rydberg Decay}

In previous work, Rydberg decay is typically accompanied by a re-initialization process, which involves either replacing leaked atoms with new atoms prepared in the state $\ket{1}$ \cite{sahay2023high} or waiting until the Rydberg population returns to the qubit subspace \cite{jandura2024surfacecodestabilizermeasurements,ma2023high}. Both of these methods incur substantial time overhead: the former requires moving atoms from a reservoir \cite{v7ny-fg31,Chiu_2025,li2025fastcontinuouscoherentatom}, and the latter introduces an idle period proportional to the lifetime of the Rydberg state.\\

Our approach differs from earlier work by directly addressing the leaked state $\ket{L}$. As previously discussed \cite{cong2022hardware}, this leaked state $\ket{L}$ includes Rydberg states $\ket{r}$ from the BBR error, and low-lying states $\ket{l}$ from the RD error. The low-lying states $\ket{l}$ refer to other hyperfine states $\ket{f}$ for alkali atoms, or ground states $\ket{g}$ in alkaline-earth atoms \cite{wu2022erasure}. The population in the Rydberg state can be efficiently converted to atom loss by applying anti-trapping light to eject atoms from the trap. Alternative rapid methods, such as laser-induced or field-induced ionization, also exist \cite{PRXQuantum.5.040343,cong2022hardware}. Additionally, a microwave-enhanced process has been demonstrated to convert $\ket{r}$ into atom loss \cite{Ebadi_2021}. In their experiment, a 100ns microwave pulse converts the Rydberg state $\ket{r=70S}$ for $\ce{^{87}_{}Rb}$ to atom loss with 98.6\% loss fidelity. The remaining includes $\pi$-pulse error (failure to transit $\ket{g}$ to $\ket{r}$) and detection error (failure to transit $\ket{r}$ to atom loss). Regarding the detection error branch, atoms lingering in the Rydberg state will decay back to the ground-state hyperfine levels. This includes a small fraction that returns to the computational subspace $|0\rangle$ and $|1\rangle$, and a majority that leaks into other hyperfine levels $|f\rangle$ outside the qubit manifold. The former is treated as a Pauli error, while the latter is similarly categorized as an RD error.\\

A critical advantage of converting population to atom loss is that lost atoms no longer participate in Rydberg interactions. This makes the error channel of Rydberg decay similar to that of amplitude damping \cite{jayashankar2022achieving}, except that the leaked state (or atom loss) can be readily distinguished from the qubit subspace through a three-outcome measurement \cite{suchara2015leakage,PRXQuantum.5.040343}. The above considerations for blackbody radiation (BBR) similarly apply to radiative decay (RD), provided the leaked state from RD does not couple to single- or two-qubit driving lasers. This requirement is naturally fulfilled in $\ce{^{171}_{}Yb}$ atoms, as most RD transitions decay into a ground state $\ket{g}$, which is energetically distinct from the qubit subspace \cite{wu2022erasure}. For alkali atoms employing hyperfine states as qubit levels, a magnetic field can be applied to separate other hyperfine states $\ket{f}$ from the qubit subspace \cite{PRXQuantum.5.040343}. 
Using this method, researchers in reference \cite{PRXQuantum.5.040343} experimentally demonstrated that the vast majority (approximately 90\%) of other hyperfine states $\ket{f}$ act similarly to an absent atom. If a higher selectivity is needed, the leaked state can be transferred to an isolated state using microwave pumping \cite{PhysRevX.13.041051}.\\

Regarding how to implement the projective measurement that distinguishes between $|L\rangle$ (including $|l\rangle$ and atom loss), $|0\rangle$, and $|1\rangle$, we discuss the protocols for alkaline-earth atoms and alkali atoms separately. For alkaline-earth atoms such as $\ce{^{171}_{}Yb}$, one can first perform a fast, non-destructive detection of the atoms that have decayed back to the ground state, as described in the erasure conversion protocol \cite{wu2022erasure}. This is followed by a projective measurement capable of discriminating among atom loss, state $|0\rangle$, and state $|1\rangle$ \cite{li2025fastcontinuouscoherentatom}. For alkali atoms such as $\ce{^{87}_{}Rb}$, The existing spin-to-position conversion already enables loss-resolved qubit readout \cite{Bluvstein_2025}. We further augment this protocol to allow for the discrimination of other hyperfine states $|f\rangle$. We first transit atoms in state $\ket{1}=\ket{F=2,m_F=0}$ into the dark state while maintaining other populations unchanged by microwave shelving pulses or Raman pulses \cite{PhysRevX.13.041051,cong2022hardware}. Then we spatially separate the dark state from other states through the spin-to-position conversion. And we then transit $\ket{0}=\ket{F=1,m_F=0}$ to $\ket{F=2,m_F=0}$ through Raman pulses and repeat the procedure above to separate $\ket{0}$ from the remaining hyperfine states $\ket{f}$. Ultimately, standard fluorescence imaging is utilized to determine the occupancy of three resolved spatial sites. An atom detected at one of these three locations corresponds to states $|0\rangle$, $|1\rangle$, or $|f\rangle$, while the absence of a signal at all sites indicates an atom loss event. \\

\subsection{The Error Channel with Pauli-Twirling and the Error Propagation}
To study the quantum operation in the presence of the leaked state, we expand the qubit system ($\ket{0},\ket{1}$) to a qutrit system ($\ket{0},\ket{1},\ket{L}$). The Pauli matrix gives $P_i' = P_i \oplus \ket{L}\bra{L}$ ($P_i = I,X,Y,Z$ is a qubit Pauli matrix). From the hardware perspective, only the state $\ket{11}$ obtains an additional phase $e^{i\pi}$ so the gate operator gives $CZ = I'\otimes I' - 2\ket{11}\bra{11}$. For the jump operator $K_1 = \sqrt{p_e}\ket{L}\bra{1}$, we prove that it propagates a Z error using forward propagation as below.
\begin{equation}
\begin{aligned}
CZ \; I'\otimes K_1 \; CZ &= \sqrt{p_e}(I'\otimes I' - 2\ket{11}\bra{11})(\ket{0L}\bra{01}+\ket{1L}\bra{11}+\ket{LL}\bra{L1})(I'\otimes I' - 2\ket{11}\bra{11})\\
&= \sqrt{p_e}(\ket{0L}\bra{01}-\ket{1L}\bra{11}+\ket{LL}\bra{L1})\\
&= \sqrt{p_e}(\ket{0}\bra{0}-\ket{1}\bra{1}+\ket{L}\bra{L})\otimes\ket{L}\bra{1} = Z'\otimes K_1
\end{aligned}
\end{equation}
The equation above naturally sets up for the jump operator $K_{1L}$. For the jump operator $K_{0L}$, we can prove that it does not propagate any error in the same way. Then we consider the condition that a leaked qubit interacts with three other qubits $i,j,k$ without leakage. If the jump operator is $K_{0L}$, it propagates to no error. Otherwise, the jump operator $K_{1L}$ propagates to a correlated Z error $Z_i Z_j Z_k$. This error propagation principle is important for our distance-preserving property.\\

For the state (or the density matrix) we consider, we assume that the density matrix can be written as the direct sum of the qubit subspace and the leakage subspace.
\begin{equation}
\rho = \rho_q \oplus \rho_L = 
\begin{bmatrix}
\rho_{00} & \rho_{01} & 0\\
\rho_{10} & \rho_{11} & 0\\
0 & 0 & \rho_{LL}
\end{bmatrix}
\end{equation}
With this equation, we can simply divide the qubit subspace and the leakage subspace when implementing the Pauli Twirling Approximation (PTA). This assumption is also easy to satisfy because both BBR and RD are decoherence processes \cite{cong2022hardware}. \\

Then we consider PTA in the presence of the leaked state. The Kraus operator and the channel of the Rydberg decay are shown below (For clarity, we consider the single-qubit channel).

\begin{equation}
\begin{cases}
K_0 = \ket{0}\bra{0} + \sqrt{1-p_e}\ket{1}\bra{1} + \ket{L}\bra{L}\\
K_1 = \sqrt{p_e}\ket{L}\bra{1}
\end{cases}
\end{equation}
To implement PTA, we add a random Pauli operator before and after the channel and take the average of the results, as below. Different from previous work, we use the Pauli matrix in the extended qutrit space.
\begin{equation}
\overline{\xi(\rho)}^{P'} = \frac{1}{4}\sum_{i=1}^4 P_i'\xi(P_i'\rho P_i') P_i' = \frac{1}{4}\sum_{i=1}^4 P_i'(\sum_{j=0,1}K_j P_i'\rho P_i'K_j^{\dagger}) P_i'
\end{equation}
We consider the no-jump evolution and the jump evolution separately by dividing the sum over \textit{j}. For no-jump evolution, the twirled channel gives ($K_{0q} = \ket{0}\bra{0} + \sqrt{1-p_e}\ket{1}\bra{1} \approx (1-\frac{p_e}{4})I+\frac{p_e}{4}Z$)
\begin{equation}
\begin{aligned}
\overline{\xi_0(\rho)}^{P'} &= \frac{1}{4}\sum_{i=1}^4 P_i'(K_0 P_i'\rho P_i'K_0^{\dagger}) P_i'\\
& =\frac{1}{4}\sum_{i=1}^4 (P_i\oplus \ket{L}\bra{L}) (K_{0q}\oplus \ket{L}\bra{L}) (P_i\oplus \ket{L}\bra{L}) (\rho_q \oplus \rho_L) (P_i\oplus \ket{L}\bra{L}) (K_{0q}^{\dagger}\oplus \ket{L}\bra{L}) (P_i\oplus \ket{L}\bra{L})\\
& = (\frac{1}{4} \sum_{i=1}^4 P_i K_{0q} P_i \rho_q P_i K_{0q}^{\dagger} P_i) \oplus \rho_L\\
& \approx [(1-\frac{p_e}{2}+\frac{p_e^2}{16})I\rho_q I + \frac{p_e^2}{16} Z\rho_q Z] \oplus \rho_L
\end{aligned}
\end{equation}
The first term represents no-jump evolution in the qubit subspace, which can be considered as a Z error with probability to the second order of $p_e$. The second term represents that the state in the leakage subspace remains unchanged. Then, the jump evolution is shown below.
\begin{equation}
\begin{aligned}
\overline{\xi_1(\rho)}^{P'} &= \frac{1}{4}\sum_{i=1}^4 P_i'(K_1 P_i'\rho P_i'K_1^{\dagger}) P_i'\\
& = \frac{p_e}{4}\sum_{i=1}^4 P_i' \ket{L}\bra{1} P_i'\rho P_i'\ket{1}\bra{L} P_i'\\
& = \frac{p_e}{4}\sum_{i=1}^4 \ket{L}\bra{1} P_i'\rho P_i'\ket{1}\bra{L}\\
& = \frac{p_e}{4} \ket{L}\bra{1} [(\sum_{i=1}^4 P_i\rho_q P_i)\oplus \rho_L]\ket{1}\bra{L}\\
& = \frac{p_e}{2}(1-\rho_{LL})\ket{L}\bra{L}
\end{aligned}
\label{A6}
\end{equation}

\begin{figure*}[h]
\renewcommand{\thefigure}{S1}
    \centering
    \includegraphics[width=0.35\textwidth]{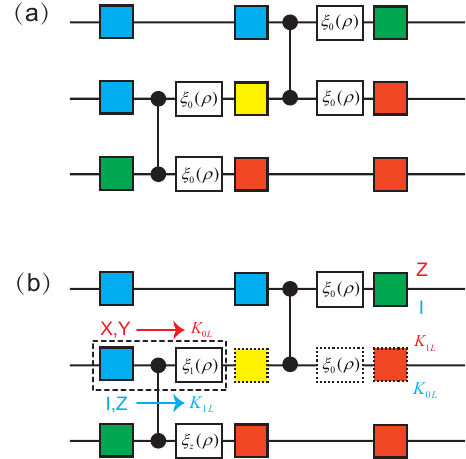}
    \caption{Pauli Twirling in the Presence of the Leaked State. Different colored squares represent different random Pauli operators in randomized compiling \cite{Wallman_2016}. White rectangles represent different channels. $\xi_0(\rho),\xi_1(\rho)$ are no-jump evolution and jump evolution defined by $K_0$ and $K_1$. $\xi_Z(\rho)$ is a tailored Z error, representing a leakage during the gate operation that totally dephases another qubit. \textbf{(a)} For the no-jump evolution, Pauli twirling cancels the non-diagonal term in the process matrix. \textbf{(b)} If the second qubit encounters a jump operator, subsequent Pauli operator, and the no-jump evolution acts trivially (dashed line). The effective jump operator is determined by the Pauli operator before. The error propagation is considered by the forward-propagation.}
    \label{figS1}
\end{figure*}

The third equation comes from the fact that $P'_i$ acts trivially on the leaked state $\ket{L}$, and the final equation comes from $\frac{I}{2} = \frac{\rho + X\rho X+Y\rho Y+Z\rho Z}{4}$ for an arbitrary density matrix in the qubit subspace. $(1-\rho_{LL})$ is the trace of $\rho_q$. Eq.\ref{A6} shows that a random Pauli matrix before the jump operator just transforms the biased-erasure channel to the erasure channel. Due to the explicit presence of the leaked state $\ket{L}$, we cannot simply treat the error as a random Pauli error, as typically done in the PTA. Instead, we start from the third line in Eq.\ref{A6}. It can be considered as a part of the channel in the operator-sum form ($K_{0L} = \sqrt{\frac{p_e}{2}}\ket{L}\bra{0}$ and $K_{1L} = \sqrt{\frac{p_e}{2}}\ket{L}\bra{1}$). 
\begin{equation}
\begin{aligned}
\overline{\xi_1(\rho)}^{P'} &=
\frac{p_e}{4}\sum_{i=1}^4 \ket{L}\bra{1} P_i'\rho P_i'\ket{1}\bra{L}\\
& = K_{0L}\rho K_{0L}^{\dagger} + K_{1L}\rho K_{1L}^{\dagger}
\end{aligned}
\end{equation}
Therefore, if one qubit encounters a leakage during a CZ gate, the probability of jump operator $K_{0L}$ or $K_{1L}$ is 50\%, respectively. This fact can also be understood by sampling a random Pauli matrix before the leaked CZ gate, Pauli $X,Y$ correspond to the jump operator $K_{0L}$, and Pauli $I,Z$ correspond to the jump operator $K_{1L}$. After the sampling, we study the error propagation according to the two jump operators $K_{0L}$ and $K_{1L}$ as discussed previously.\\

In our main text, we only consider CZ gates to generate the RHG cluster state. 
The jump operators $K_{0L}$ and $K_{1L}$ exhibit an error propagation behavior similar to Pauli errors under the CZ gate; however, this property does not hold generally for leakage errors. For a jump operator $K_{1L}$ in the target qubit of the CNOT gate $CNOT = (I'\otimes I' - \ket{11}\bra{11} - \ket{10}\bra{10} + \ket{10}\bra{11} + \ket{11}\bra{10})$, the forward propagation gives 

\begin{equation}
\begin{aligned}
CNOT \; I'\otimes K_{1L} \; CNOT &= \sqrt{p_e/2} \; CNOT(\ket{0L}\bra{01}+\ket{1L}\bra{11}+\ket{LL}\bra{L1})CNOT\\
&= \sqrt{p_e/2}(\ket{0L}\bra{01}+\ket{1L}\bra{10}+\ket{LL}\bra{L1})\\
&= \sqrt{p_e/2}(\ket{0}\bra{0}+\ket{L}\bra{L})\otimes\ket{L}\bra{1} +\ket{1}\bra{1}\otimes\ket{L}\bra{0}\\
&= \frac{1}{2}(I+Z')\otimes K_{1L}+\frac{1}{2}(I-Z')\otimes K_{0L}
\end{aligned}
\end{equation}

Namely, an error in the target of the CNOT gate represented by the jump operator $K_{1L}$ will induce both $K_{1L}$ and $K_{0L}$. This is a problem encountered when dealing with the amplitude damping error \cite{jayashankar2022achieving}. In our condition, we can utilize the freedom in the operator-sum form of the channel to define a set of Kraus operator that is \textit{noise-structure preserving} in the CNOT gate, as shown in the equation below ($K_{+L} = \sqrt{\frac{p_e}{2}}\ket{L}\bra{+}$ and $K_{-L} = \sqrt{\frac{p_e}{2}}\ket{L}\bra{-}$). It is easy to check that $K_{-L}$ in the target qubit propagates a Z error to the control qubit while $K_{+L}$ in the target qubit does not propagate an error. However, despite this transformation existing, we notice that the Kraus operator is not \textit{noise-structure preserving} if the control and target of the CNOT gates are exchanged. See our follow-up study for this condition \cite{zchg-x177}.

\begin{equation}
\begin{aligned}
\overline{\xi_1(\rho)}^{P'} &=
\frac{p_e}{4}\sum_{i=1}^4 \ket{L}\bra{1} P_i'\rho P_i'\ket{1}\bra{L}\\
& = K_{0L}\rho K_{0L}^{\dagger} + K_{1L}\rho K_{1L}^{\dagger}\\
& = K_{+L}\rho K_{+L}^{\dagger} + K_{-L}\rho K_{-L}^{\dagger}
\end{aligned}
\end{equation}
To some conclusions, after Pauli-twirling, the initial channel studied in the biased erasure error is just the erasure error now (by omitting no-jump evolution, which is the Z error to second order). And the jump operators (errors) propagate to subsequent CZ gates in a similar way to the Pauli error. Our work also applies to a larger range of leakage errors as long as the leaked state $\ket{L}$ is not involved, although the initial proposal of this work is to deal with the Rydberg decay. For example, the atom loss error is modeled as a biased erasure error if it results from the conversion of BBR error, and is modeled as an unbiased erasure error if it results from coherent movement \cite{Baranes_2026,Perrin_2025}. Our scheme is compatible with both conditions because the Pauli twirling procedure results in an identical unbiased channel, irrespective of whether the original noise channel was biased or unbiased. The picture is derived as Fig.\ref{figS1}.

\section{Structure of the RHG Cluster State and Gate Sequence}\label{appB}
\begin{figure*}[h]
\renewcommand{\thefigure}{S2}
    \centering
    \includegraphics[width=0.8\textwidth]{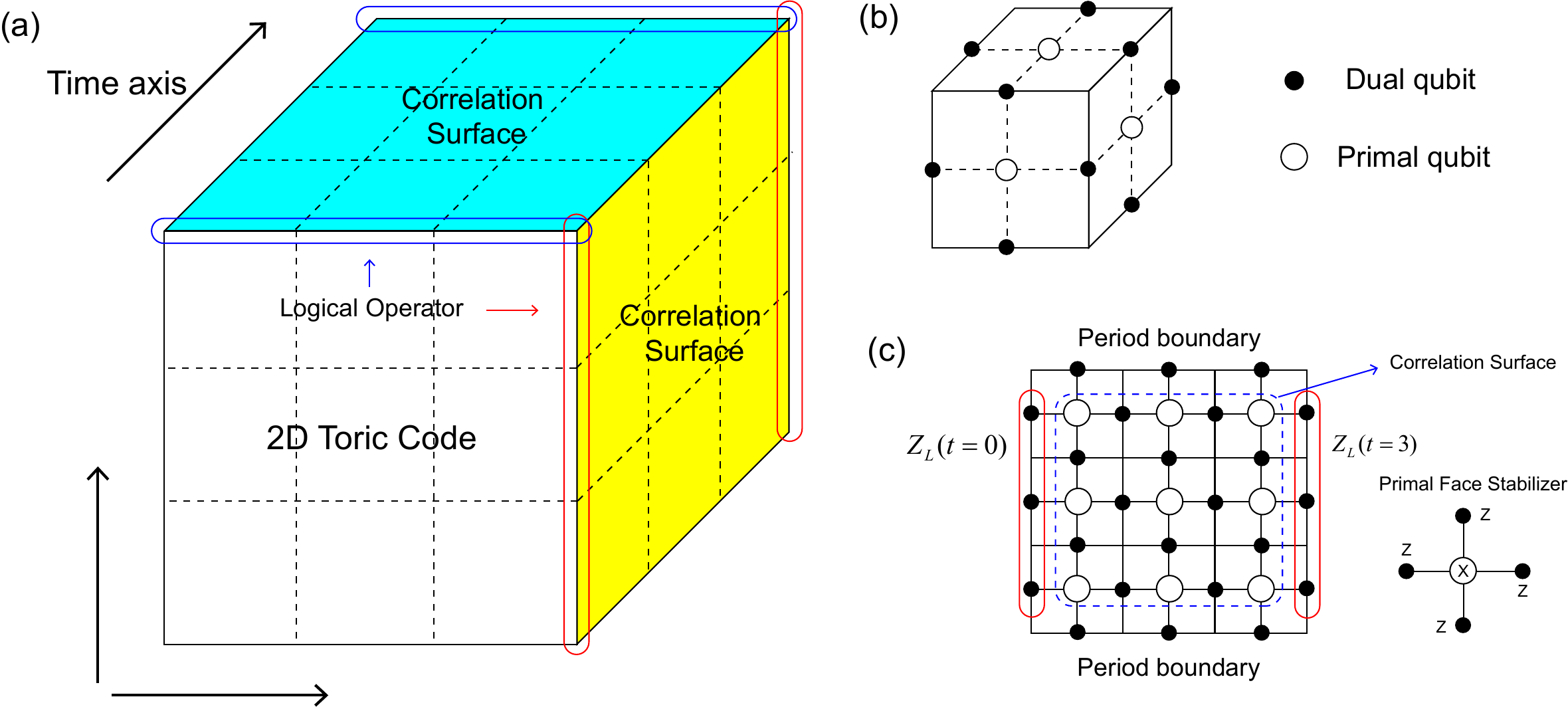}
    \caption{The structure of the RHG cluster state. \textbf{(a)} A RHG cluster state as a toric code propagating in time. The dashed lines divide the whole cubic into $3\times3\times3$ unit cells, whose structure is shown in (b). Logical operators and corresponding correlation surfaces are marked with a rounded rectangle and colored face, respectively. \textbf{(b)} A unit cell in the RHG cluster state. Primal qubits lie on the face while dual qubits lie on the edges (in the primal lattice). \textbf{(c)} Primal correlation surfaces connect logical operators defined on dual qubits in different time series. By multiplying the primal face stabilizers on the correlation surface, we map $Z_L(t=0)$ to $(-1)^k*Z_L(t=3)$, where $k$ is the parity of the X measurement result on the correlation surface.}
    \label{figS2}
\end{figure*}

In this section, we elaborate on the structure of the RHG cluster state explicitly and connect it to the view that the RHG cluster state can be regarded as a toric code propagating in time. See Fig.\ref{figS2}, we plot a $3\times3\times3$ RHG cluster state, and we can select two directions to define the toric code (with period boundary conditions). Another direction is the time axis. The plotted lattice is defined as the primal lattice. One can also define the dual lattice by placing dual qubits on the face of the unit cell \cite{fowler2009topologicalclusterstatequantum}. This definition transitions the primal cell to the dual vertex; the primal face to the dual edge; the primal edge to the dual face; and the primal vertex to the dual cell, respectively. Multiplying six face stabilizers defined in a unit cell gives the cell X stabilizer, which detects the Z error in decoding and error correction. A logical operator consists of dual qubits propagates in time by multiplying primal face stabilizers, and whether a logical error has happened is determined by checking the parity of the primal correlation surface. The dual lattice is considered in the same way, and the primal lattice and the dual lattice are decoded independently \cite{fowler2009topologicalclusterstatequantum}.\\

\begin{figure*}[h]
\renewcommand{\thefigure}{S3}
    \centering
    \includegraphics[width=0.8\textwidth]{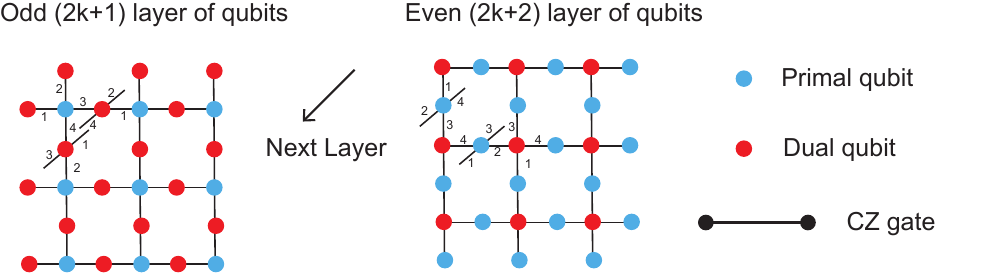}
    \caption{One CZ gate sequence that arranges $Z_3 Z_4$ on different directions. Note that, for the first and the last layers, some CZ gates do not exist.}
    \label{figS3}
\end{figure*}

Having clarified the relationship between the correlation surface and the logical error, we need to guarantee that the hook error $Z_3 Z_4$ does not occur along the same direction. One possible gate sequence is shown in Fig. \ref{figS3}.
\section{Details in Numerical Simulation}\label{appC}
In this section, we provide some details about our simulation. The details include basic settings, some extended data, and the reasonableness of ignoring the correlated leakage. 
\subsection{Basic Settings}
In our simulation, we account for the physical error during CZ gates. In addition to Rydberg decay errors, we also consider two-qubit depolarization errors for generality. We set $p = p_e + p_d$, where $p_e$ refers to the Rydberg decay error rate in the main text while $p_d$ refers to the two-qubit depolarization error rate. We simulate a $d\times d\times d$ RHG lattice with period boundary conditions in two directions. In the third direction (time axis), there are $2d+1$ layers of qubits. The CZ gates in the first layer and the last layer are noiseless, because we assume that state-initialization and final measurement are ideal, similar to circuit-based error correction simulation \cite{Baranes_2026}. We derive the logical error by decoding the primal lattice, accounting for the parity of a single correlation surface.\\

The logical error rate with deviation is estimated through \textit{proportion\_confint} function \cite{seabold-proc-scipy-2010} in \textit{statsmodels.stats.proportion}, by calculating $99\%$ confidence interval. The majority of the time consumption is constructing the decoding graph through the sampled leakage instance in CZ gates, instead of the decoding itself. Therefore, we apply the numerical technique that we only generate a certain amount of leakage instances (no less than $1.6\times10^4$ to guarantee accuracy) \cite{PhysRevResearch.7.013249}, when considering a mixture of the Rydberg decay error and the two-qubit depolarization error. A reduced number of leakage instances may compromise the accuracy of the estimated logical error rate and inflate the variance across independent trials. In this work, we determined the sufficiency of our chosen number by heuristically increasing the value and monitoring whether the logical error rate and its variance exhibited any statistically distinguishable changes. With the current parameter settings, the construction of decoding graphs does not constitute a major computational overhead. However, we anticipate that for future simulations involving higher $R_e$ ($R_e > 0.9$ or even $>0.95$) — and potentially for larger code distances — an increased number of leakage instances will be required. In such scenarios, decoding graph construction is expected to emerge as the primary bottleneck impeding runtime optimization. The threshold for the pure Rydberg decay is estimated by fitting the universal scaling ansatz for the critical point of phase transition \cite{WANG200331} with $d = \{7,9,11,13\}$ and an error range from $0.34$ to $0.38$, while the threshold shown in Fig \ref{figS4} is derived from the intersection point of $d = 9$ and $d = 11$ \cite{wu2022erasure}.\\

To formulate the error model for Rydberg decay, we adapt the error model from \cite{sahay2023high}. Our model accounts for noisy two-qubit gates (CZ gates) with Rydberg decay error rate $p_e$. When neither of the two qubits is leaked before the CZ gate, the CZ errors are represented by $\frac{p_e}{4}\{L\otimes I,L\otimes Z,I\otimes L,Z\otimes L\}$. We write the leakage error $L$ explicitly because the leaked atoms exist until the final measurement. Here, we ignore the correlated leakage in Rydberg decay by arguing that it is less harmful than the single-leakage error we considered here (See \ref{appC} 3.The Reasonability to Ignore the Correlated Leakage) \cite{wu2022erasure}. Once a leakage error happens on the qubit, we assign it to be the jump operator $K_{0L}$ or $K_{1L}$ with $50\%$ probability. The error propagation is considered by forward-propagation, and we prove that $K_{0L}$ propagates to no error and $K_{1L}$ propagates to a correlated Z error through successive CZ gates. (See \ref{appA} for details))\\

The two-qubit depolarization channel is described by $\frac{p_d}{15}\{I,X,Y,Z\}^{\otimes 2}\setminus{\{I\otimes I\}}$ for each CZ gate. In the RHG cluster state, the primal lattice and the dual lattice are decoded separately. For the primal lattice, stabilizers are products of the X operator of face qubits, so only Z errors flip the stabilizers. X errors in dual qubits propagate to correlated Z errors in primal qubits. Therefore, we only need to consider three types of error in the two-qubit depolarization error model, namely $\{Z_p,Y_p\} \otimes \{I_d,Z_d\}$, $\{I_p,X_p\} \otimes\{X_d,Y_d\}$ and $\{Z_p,Y_p\} \otimes \{X_d,Y_d\}$, each type with probability $\frac{4 p_d}{15}$ (The subscript of Pauli matrix refers to the primal lattice or the dual lattice). After accounting for all three types of errors, each edge representing the primal qubit has $\frac{32p_d}{15}$ error probability and primal hook error $Z_3 Z_4$ relative to each dual qubit has $\frac{8p_d}{15}$ error probability, except for the primal qubits on the $2^{nd}$ and $(2d-1)^{th}$ layers.  Primal qubits on the $2^{nd}$ and $(2d-1)^{th}$ layers have no hook error and have single qubit error $\frac{28 p_d}{15}$ because we have set the CZ gates in the first layer and the last layer noiseless.\\

We also include a simulation on the biased erasure in our article \cite{sahay2023high}, which is simple to simulate in the RHG cluster state. The biased erasure channel commutes with all CZ gates, so we can put it at the end of the circuit without propagating any error. Namely, each primal qubit has $1-(1-p_e)^4$ probability of having the biased erasure. The two-qubit depolarization error in biased erasure conversion with the RHG cluster state is considered in the same way as our protocol, as described in the last paragraph.\\

\subsection{Extended Discussion and Data for the comparison with biased erasure conversion}
In the main text, we choose to compare our method with biased erasure conversion (BE) on the RHG cluster state. This comparison serves as a representative and fair benchmark. However, we also acknowledge that biased erasure conversion is a versatile technique applicable to other error correction codes, such as the XZZX code, or measurement-based protocols, such as hybrid fusion requiring post-selection \cite{sahay2023high}. Therefore, in \ref{appE}, we present another comparison with (un)biased erasure conversion + XZZX code to illustrate the respective resource overheads. In this section, we provide two sets of extended data as a supplement to the main text and extended discussion for the comparison with BE on the RHG cluster states.\\

We first compare the error models for BE and our protocol. For each CZ gate, if a leakage happens, we detect the leakage and replace the leaked atom with a new atom in $\ket{1}$, which results in an error channel $\frac{p_e}{4}\{I\otimes I,I\otimes Z,Z\otimes I,Z\otimes Z\}$, namely two $50\%\;Z$ errors on the two involved qubits. Note that these errors commute with the CZ gates used for RHG state preparation; consequently, they do not induce extra error propagation. For our protocol, the CZ errors are represented by $\frac{p_e}{4}\{L\otimes I,L\otimes Z,I\otimes L,Z\otimes L\}$, which introduces a $50\%\;Z$ error on one qubit and a leakage error on another qubit. The leakage error propagates to correlated Z errors through subsequent CZ gates, as shown in the main text, Fig.2(b). In other words, in contrast to the biased erasure method, which employs mid-circuit detection and atom replacement, our approach introduces more propagated errors. The comparison between these two methods, therefore, highlights the impact of such propagated errors.\\

\begin{figure*}[h]
\renewcommand{\thefigure}{S4}
    \centering
    \includegraphics[width=0.6\textwidth]{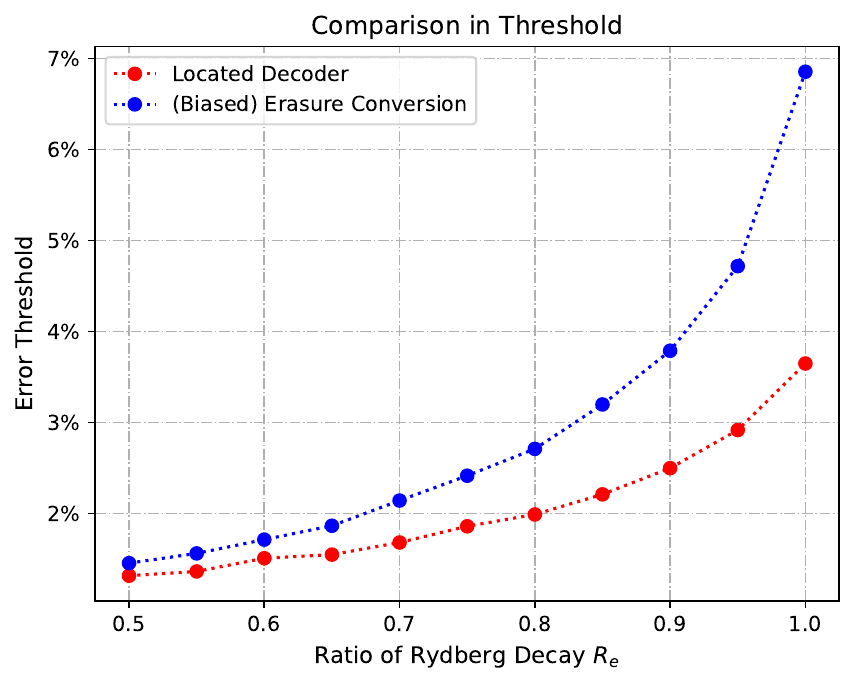}
    \caption{Comparison between biased erasure and located decoder on the threshold: Biased erasure conversion always has a higher threshold, and the advantage reduces when $R_e$ declines. The threshold points are derived from intersection points of $d=9$ and $d=11$, and each logical error rate point is sampled for $10^5$ with $2\times10^4$ leakage samples.}
    \label{figS4}
\end{figure*}

We then present a threshold comparison between our method and BE, as shown in Fig. \ref{figS4}. Unsurprisingly, our method yields a lower threshold compared to BE, which is attributed to the contribution of additional propagation errors. Note that the threshold of BE for $R_e=1$ is $6.854\%$. We can verify that the total erasure rate $1-(1-p_e)^4=0.2472\approx 0.249$, which is the percolation threshold on a 3D cubic lattice \cite{PhysRevLett.105.200502}.\\

The effective error distance $d_e$ is a physical quantity that characterizes the scaling behavior of the logical error rate with respect to the physical error rate, satisfying $p_L\sim (\frac{p}{p_{th}})^{d_e}$ \cite{wu2022erasure,d1v7-nctj}. For one type of error, this quantity is connected to the number of errors the code corrects. For example, a quantum code with an odd code distance $d$ corrects $\frac{d-1}{2}$ Pauli errors, so the effective error distance for Pauli errors is $d_e = \frac{d+1}{2}$, representing $d_e$ Pauli errors are needed to introduce a logical error when the physical error rate is small. The code also corrects $d-1$ located errors, thus the effective error distance for located errors is $d_e = d$ \cite{Nielsen_Chuang_2010}. In the presence of both Pauli and located errors, logical errors stem from both individual error types and their combinations. Therefore, in this scenario, the effective error distance satisfies $d>d_e>\frac{d+1}{2}$ \cite{wu2022erasure}. Crucially, the value of the effective error distance varies across different physical error rate regimes, reflecting the shifting roles of the two error types. When the physical error rate is near the threshold, and the proportion of located errors is high, located errors make a significant contribution to the logical error rate, resulting in a $d_e$ substantially larger than $(d+1)/2$. Conversely, at sufficiently low physical error rates, the logical error rate is dominated by the harder-to-correct Pauli errors, causing $d_e$ to approach $(d+1)/2$ \cite{wu2022erasure}. This suggests that a higher effective error distance cannot be extrapolated to imply lower logical error rates at low physical error probabilities in the presence of both error types. This is because, as the physical error rate decreases, the effective error distance gradually converges to $d_e = \frac{d+1}{2}$, causing the advantage in effective error distance to gradually vanish.\\

\begin{figure*}[h]
\renewcommand{\thefigure}{S5}
    \centering
    \includegraphics[width=\textwidth]{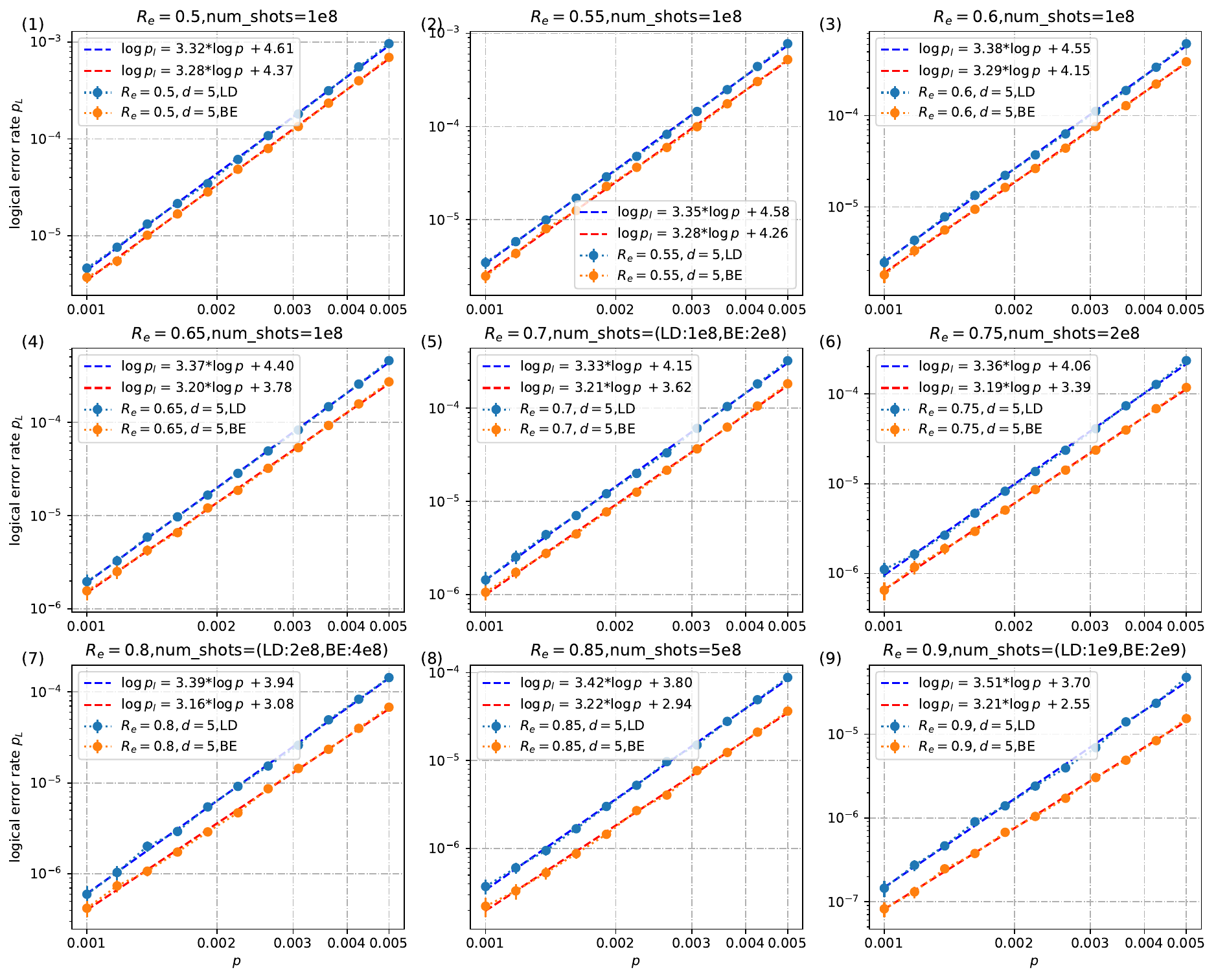}
    \caption{Initial data in Fig.4(a): Each point for logical error rate samples for $1.6\times 10^4$ leakage instance, and the number of total shots is adjusted to ensure the logical failure shots are larger than 100.}
    \label{figS5}
\end{figure*}

In our study, we selected an \textit{identical physical error rate interval} $0.1\%$ to $0.5\%$ to derive the effective error distance for both methods, ensuring a fair comparison. More importantly, in this regime, the effective error distance carries a clear physical interpretation: In our decoding strategy, these additional propagated errors are decoded as located errors. Consequently, given identical Pauli error rates, our scheme exhibits a higher proportion of located errors, which results in a larger effective error distance. The key distinction of our method compared to BE lies in the fact that, although we permit additional propagation errors for the sake of hardware simplicity, we explicitly transform them into located errors during decoding. This ensures that the resulting compromise in logical error performance remains minimal. The initial data to derive the error distance shown in Fig.4(a) are provided in Fig.\ref{figS5}. It seems strange that the effective error distance does not increase monotonically with $R_e$, as shown previously in \cite{wu2022erasure}. This is because we have \textit{fixed the physical error rate range} for both BE and LD, instead of \textit{varying the range according to the threshold}. Therefore, when $R_e$ increases, the threshold is larger, but the error range for deriving the effective error distance is not changed, which leads to a probable decay in the effective error distance \cite{wu2022erasure}.\\

We emphasize that demonstrating a larger effective error distance plays a pivotal role in addressing concerns regarding our scheme's lower threshold and in identifying its specific advantage regime. We can illustrate this point using a proof by contradiction: Without a larger effective error distance, we may estimate the logical error rate for low physical error rates by $p_{L,BE}=C_1(\frac{p}{p^{BE}_{th}})^{\alpha d}$ and $p_{L,LD}=C_2(\frac{p}{p^{LD}_{th}})^{\alpha d}$ and we use the same $\alpha$ to account for the condition where our scheme can't demonstrate a higher effective error distance \cite{985g-58gd}. Then we have $\eta = \frac{p_{L,LD}}{p_{L,BE}} = \frac{C_2}{C_1}*(\frac{p^{BE}_{th}}{p^{LD}_{th}})^{\alpha d}$, namely the gap in logical error rates between two methods scales exponentially with increasing $d$. Only by demonstrating a larger effective error distance does the gap in logical error rates between the two methods narrow as the physical error rate decreases, thereby resulting in comparable performance at lower error rates.\\

\subsection{The Reasonability to Ignore the Correlated Leakage}
We have noticed that there is \textit{correlated leakage} in Rydberg decay \cite{wu2022erasure}. But we have ignored its effect in our simulation. This is reasonable because the correlated leakage is less harmful than a single leakage error $L \otimes 50\%\  Z$ and it takes up a small portion of the error ($\sim \frac{1}{10}$), as explained in detail below.\\ 

When implementing the two-qubit gate with neutral atoms, an RD error during the CZ gate leads to failure of Rydberg blockade, so it triggers another BBR error with $O(1)$ probability. Therefore, there is $O(p)$ probability error that the interacting two qubits are leaked in a single CZ gate, instead of $O(p^2)$, although the probability is small. However, it is a less harmful error in the simulation of the RHG cluster state compared to a single leakage error, provided that the generation circuit is shallow and the error propagation is limited. When decoding the primal lattice, the correlated leakage $L\otimes L$ propagates to the same error with $L(dual) \otimes 50\%\  Z(primal)$ since the leaked qubit is regarded as a completely mixed state with $50\%\ Z$ error. The difference lies in the fact that the former has a directly detected leakage in the primal qubits, while the latter only has propagated errors. The directly detected leakage is easier to decode compared to the propagated error. Therefore, the correlated leakage is less harmful than a single leakage error $L \otimes 50\%\ Z$. For simplicity and clarity, we don't consider this complex but indeed less harmful error in the simulation. 

\section{Logical Error Rate Comparison in Small Error Rate Limit and Increasing distance}\label{appD}

In this section, we give an extended discussion on the sub-threshold scaling shown in the main text. In Fig.4, we show that the error distance of our method (LD) is slightly larger than that of biased erasure conversion and demonstrate comparable performance in the logical error rate for a gate error range of interest. Furthermore, we are also interested in the performance when the physical error rate is small enough, and the distance is increasing, since it is the general method to improve the accuracy of error correction.\\

The above discussion assumes an ideal most-likely error decoder, without specifying its particular implementation. The statement is based on the \textit{phenomenological error model} \cite{Watson_2014}, but since we have arranged the hook error $Z_3 Z_4$ in different directions and our numerical results have shown that the error distance does not degrade under the \textit{circuit-level error}, it applies to our conditions safely.\\

When $p$ is small enough, the logical error rate is governed by the logical error with the shortest error chain \cite{Watson_2014}. For a code with odd distance $d$, $d_e = \frac{d+1}{2}$ Pauli errors trigger a logical error. When the \textit{located error} is present, one located error together with $\frac{d-1}{2}$ Pauli errors is enough to introduce a logical error. In our strategy, the located error comes from directly detected leakage and propagated errors from dual qubits, while in the biased erasure conversion the located error only comes from directly detected leakage (Although this argument just explains the logical error in the RHG cluster state we considered, biased erasure conversion will always have fewer types of located error in general scenarios because our method has allowed more error propagation). Now we write the logical error rate for LD and BE as the equations below (here, $p = p_e + p_d$ and $R_e$ are absorbed in the combination coefficient $C_{Pauli}(d)$, $C_{1DL}(d)$ (`1DL' means there is a single detected leakage in the error chain) and $C_{1P}(d)$ (`1P' means there is a single propagated error from a dual qubit in the error chain)).

\begin{equation}
\begin{cases}
p_{L,BE} = C_{Pauli}(d) p^{\frac{d+1}{2}} + C_{1DL}(d) p^{\frac{d+1}{2}} + o(p^{\frac{d+1}{2}})\\
p_{L,LD} = C_{Pauli}(d) p^{\frac{d+1}{2}} + C_{1DL}(d) p^{\frac{d+1}{2}} + C_{1P}(d) p^{\frac{d+1}{2}} + o(p^{\frac{d+1}{2}})
\end{cases}
\end{equation}

The ratio of the logical error rate is $\eta = \frac{p_{L,LD}}{p_{L,BE}}= 1 + \frac{C_{1P}(d)}{C_{Pauli}(d) + C_{1DL}(d)}$ we then prove two things: 1. $\frac{C_{1P}(d)}{C_{1DL}(d)} \sim O(1)$ 2. $\frac{C_{1P}(d)}{C_{Pauli}(d)} \sim O(d)$.\\

\noindent
\textit{Proof}:
To introduce a logical error, we need to sum over all possible shortest logical error chains whose length is $\frac{d+1}{2}$. For each selected error chain, it contributes to a logical failure rate $\sim (1-R_e)^{\frac{d+1}{2}} p^{\frac{d+1}{2}}$ for pure Pauli error. But for an error chain with one located error, we first need to select a location to place the located error, which has $\frac{d+1}{2}$ options. For the error chain with one directly detected leakage, it contributes to a logical failure rate $\sim \frac{d+1}{2} (1-R_e)^{\frac{d-1}{2}} R_e p^{\frac{d+1}{2}}$. For the error chain with one propagated error, it contributes to a logical failure rate $\sim C\frac{d+1}{2} (1-R_e)^{\frac{d-1}{2}} R_e p^{\frac{d+1}{2}}$ ($C$ is a constant determined by the number proximal qubits and the probability that a leaked dual qubit propagates to located errors in the primal lattice. We argue above for each selected error chain, so the conclusion is preserved when summing over all possible error chains.\\

Given the discussion above, the ratio of the logical error rate increases when $d$ increases, but it converges to a constant for large $d$.

\section{Comparison with (Biased) Erasure Conversion in XZZX codes}\label{appE}
For the sake of fairness and physical intuition, the main text benchmarks our method against Biased Erasure conversion (BE) on the RHG cluster state. However, practically speaking, erasure conversion inherently includes atom qubit updates, thereby obviating the need for the qubit updates intrinsic to MBQC. Under hardware conditions that support erasure conversion, directly implementing it on the XZZX surface code offers a pathway to lower resource overhead \cite{wu2022erasure,sahay2023high}. Accordingly, this section provides a direct comparison between implementing our protocol on the RHG cluster state versus applying erasure conversion to the XZZX code, including resource estimates for achieving a logical error rate of $10^{-12}$.\\

Here, we provide an additional distinction between biased \cite{sahay2023high} and unbiased erasure \cite{wu2022erasure}. Our protocol utilizes Pauli twirling to remove potential coherent errors, a process that effectively symmetrizes the error channel. As a result, whether the erasure is biased or unbiased does not affect our scheme. In the main text, we benchmark against the superior Biased Erasure Conversion (BE) for brevity. However, biased erasure is predicated on the condition that only atoms in the $|1\rangle$ state are susceptible to leakage and subsequent detection \cite{sahay2023high}. This assumption fails in scenarios where erasure errors are dominated by atom loss resulting from coherent movement and atom heating \cite{Baranes_2026,Perrin_2025}. Therefore, in this section, we distinguish between biased and unbiased erasure to allow for a more reasonable comparison, accounting for regimes above. We consider the limiting cases of fully biased ($\eta_e = \frac{1}{2}$) and fully unbiased ($\eta_e = \infty$) erasure, as defined in \cite{sahay2023high}. More comprehensive scenarios with $\frac{1}{2}<\eta_e<\infty$ will be considered in future work.\\

For simulation settings, we have chosen the same settings as those in \cite{wu2022erasure,sahay2023high}. Our simulations account for erasure errors and Pauli errors in two-qubit gates. We define the total error rate $p:=p_e+p_d$, with an erasure error rate $p_e$ and a Pauli error rate $p_d$, respectively. We simulate the two-qubit CX gate with a two-qubit CZ gate and two single-qubit Hadamard gates. For each CZ gate, it has a $p_e$ probability to have an erasure error, which is modeled in the simulations by applying a Pauli error chosen uniformly at random from $\{I,X,Y,Z\}^{\otimes2}$ for unbiased erasure or from $\{I,Z\}^{\otimes2}$ for biased erasure. If no erasure error happens, the CZ gate encounters a two-qubit depolarization error, represented by $\frac{p_d}{15}\{I,X,Y,Z\}^{\otimes 2}\setminus{\{I\otimes I\}}$. The ratio of erasure $R_e$ is defined as $R_e:=\frac{p_e}{p}$. For an XZZX surface code with code distance $d$, we first initialize all qubits in $\ket{+}$ and then implement a round of noiseless syndrome measurement, followed by $d$ rounds of syndrome measurement and one final round of noiseless measurement. The circuit is generated in Stim \cite{Gidney_2021}, and we decode the syndromes and derive the logical error rates using PyMatching \cite{Higgott2025sparseblossom}.\\

\begin{figure*}[h]
\renewcommand{\thefigure}{S6}
    \centering
    \includegraphics[width=0.6\textwidth]{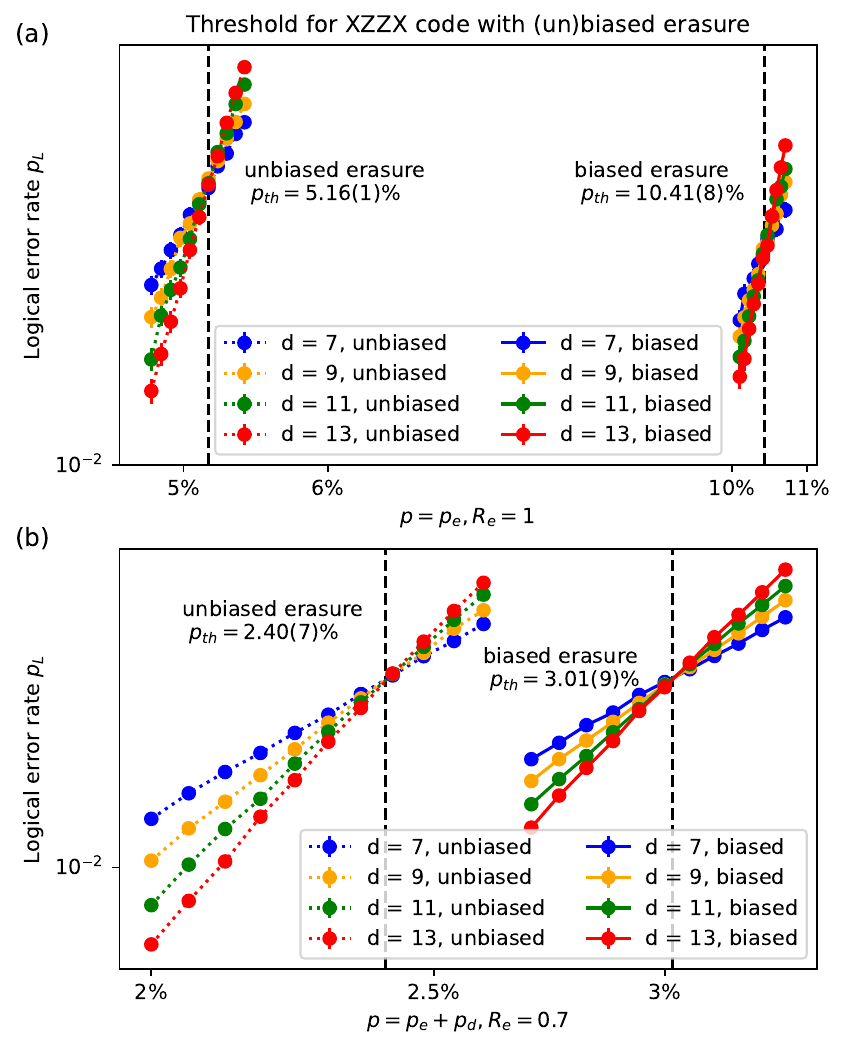}
    \caption{Thresholds for biased erasure and unbiased erasure with XZZX surface codes, for (a) $R_e=1$ and (b) $R_e=0.7$.}
    \label{figS6}
\end{figure*}

We first present the thresholds of the XZZX surface code under both unbiased and biased erasure error models, specifically for the cases of $R_e=1$ and $R_e=0.7$, in Fig.\ref{figS6}. The $R_e=0.7$ thresholds are used for subsequent fitting. The thresholds are derived by fitting the universal scaling ansatz for the critical point of phase transition \cite{WANG200331} with logical error rates demonstrated in Fig.\ref{figS6}. Our threshold results for $R_e=1$ are consistent with results in \cite{wu2022erasure,sahay2023high}, and our threshold results for $R_e=0.7$ are consistent with results in \cite{Baranes_2026} Fig S16. We infer that these small deviations likely stem from variations in the threshold extraction protocols and the distinct ranges of physical error rates sampled in different works. \\

\begin{figure*}[h]
\renewcommand{\thefigure}{S7}
    \centering
    \includegraphics[width=\textwidth]{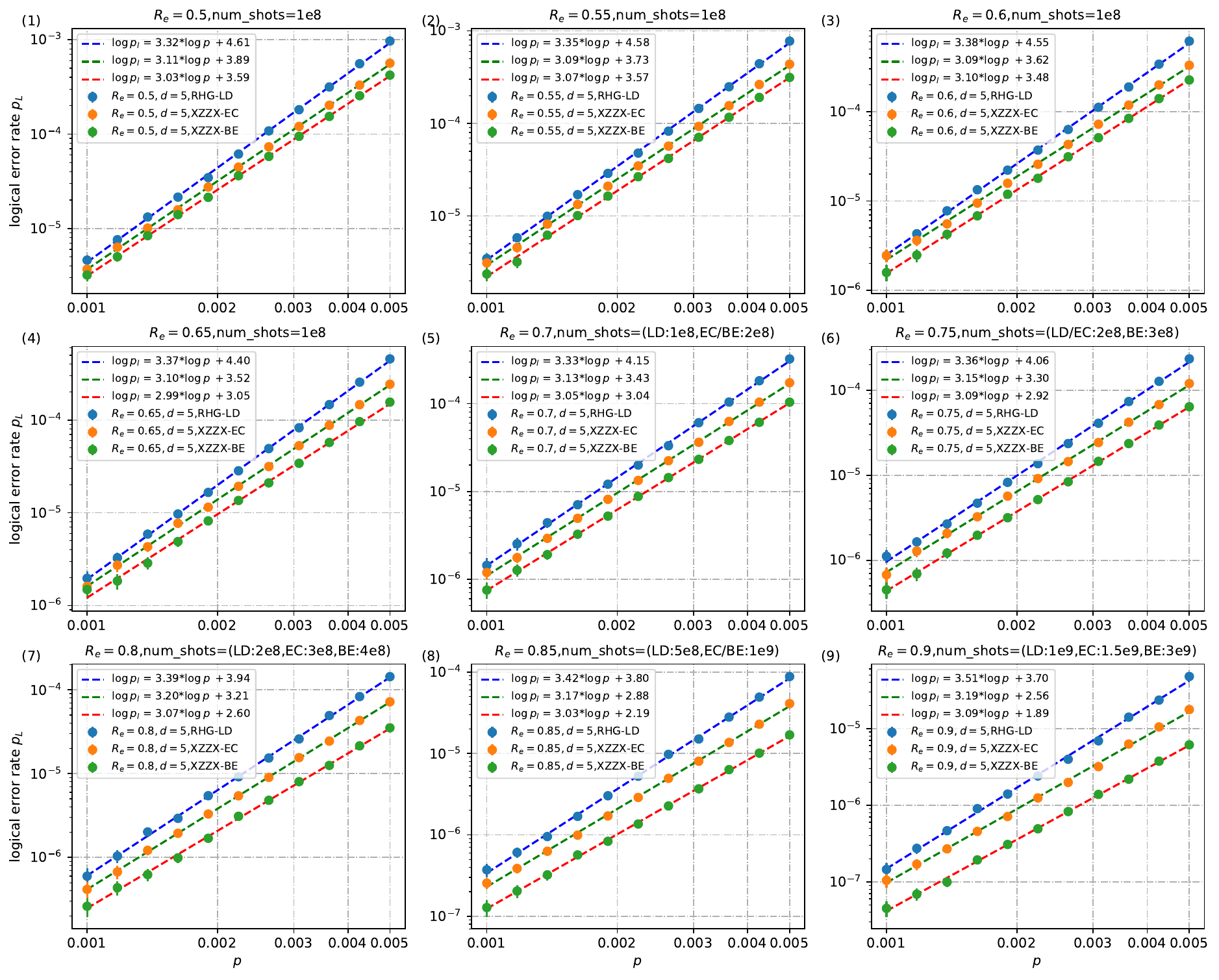}
    \caption{Sub-threshold performance for three methods. The settings are the same with results in Fig. \ref{figS5}.}
    \label{figS7}
\end{figure*}

Next, we present the performance of the three methods in the sub-threshold regime (physical error rates between $0.1\%$ and $0.5\%$) in Fig.\ref{figS7}. We denote the RHG cluster state with the located decoder as RHG-LD, and refer to the XZZX surface code with unbiased and biased erasure as XZZX-EC and XZZX-BE, respectively. The results reveal that the logical error rates satisfy $p_L^{RHG-LD}>p_L^{XZZX-EC}>p_L^{XZZX-BE}$, which are consistent with the results in Fig.\ref{figS6} satisfying $p_{th}^{RHG-LD}<p_{th}^{XZZX-EC}<p_{th}^{XZZX-BE}$. Importantly, although it lacks a clear physical counterpart, the advantage of the RHG-LD method in terms of effective error distance persists. This advantage elucidates the process by which the gap in logical error rates between RHG-LD and the other two methods narrows as the physical error rate decreases. In the low error rate regime, the RHG-LD method achieves logical error rates comparable to those of the XZZX-BE method. For $R_e=0.7,d=5,p=0.1\%$, the ratio $\eta = \frac{p_L^{RHG-LD}}{p_L^{XZZX-BE}}=\frac{1.44\times10^{-6}}{7.53\times10^{-7}}\approx1.91$. These results reinforce the comparison between RHG-LD and RHG-BE presented in the main text, further validating our core conclusion that the logical error rates are comparable in the low error rate regime.\\

\begin{figure*}[h]
\renewcommand{\thefigure}{S8}
    \centering
    \includegraphics[width=\textwidth]{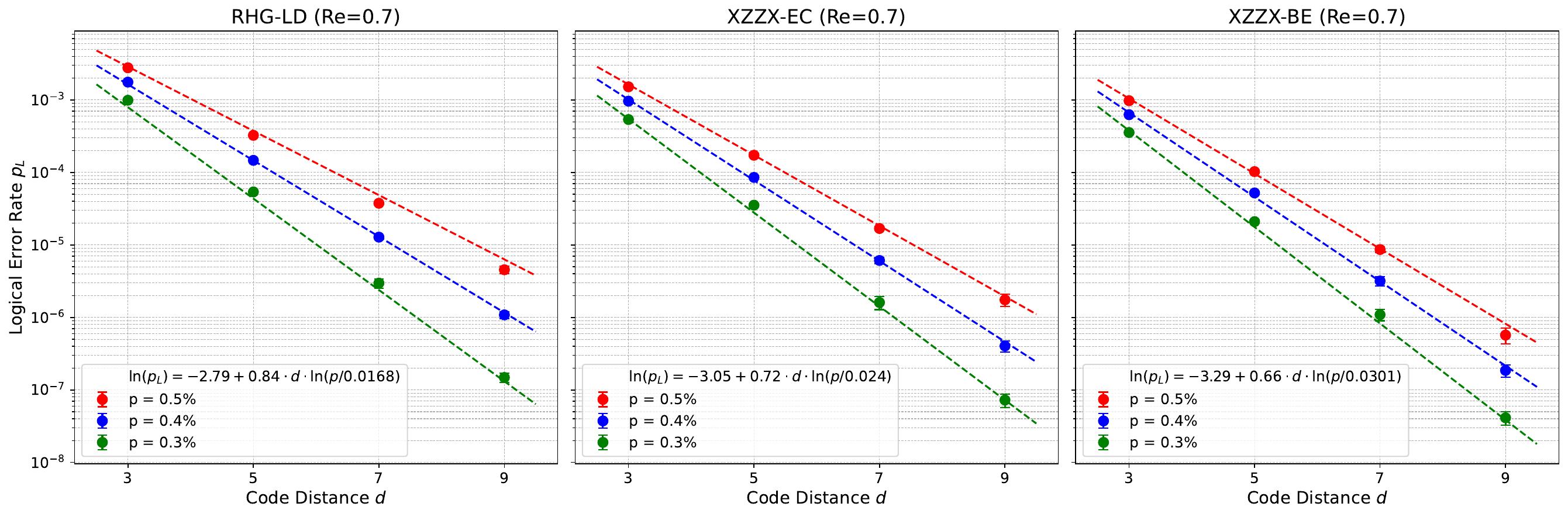}
    \caption{Logical error rate results for the ansatz $p_L=\alpha(p/p_{\text{th}})^{\beta d}$. Each point for logical error rate samples for $1.6 \times 10^4$ leakage instances, and the number of total shots is at least $10^{8}$ and is adjusted to ensure the logical failure shots are larger than 100.}
    \label{figS8}
\end{figure*}

Finally, we employ the ansatz $p_L=\alpha(p/p_{th})^{\beta d}$ (or equivalently $\ln p_L = \ln \alpha + \beta d \ln(p/p_{th})$) to fit the logical error rates calculated for code distances $d \in \{3, 5, 7, 9\}$ and physical error rates $p \in \{0.3\%, 0.4\%, 0.5\%\}$ with $R_e=0.7$ \cite{Baranes_2026,985g-58gd}. The results for the three methods are shown in Fig.\ref{figS8}. The thresholds for the three methods were pre-calculated. We then utilized \textit{numpy.polyfit} to perform the fitting and extract the parameters $\alpha$ and $\beta$. Based on the derived scaling relationship, we then estimate the resource overhead required to suppress the logical error rate to $10^{-12}$. The results are shown in the Table.\ref{tab1}. Note that the RHG cluster state with periodic boundary conditions can be regarded as the space-time propagation of a toric code, thereby encoding two logical qubits. In contrast, the XZZX surface code with open boundary conditions encodes only one logical qubit. In this table, we treat the RHG cluster state as encoding a single logical qubit to compensate for the disparity in overhead arising from the different boundary conditions employed in the two approaches.\\

\begin{table}[htbp]
    \centering
    \caption{Overhead estimation and comparison between three methods for $R_e=0.7$. We first employ the ansatz to calculate the theoretical code distance $d_{fit}$ required for $p_L=10^{-12}$, and then identify the smallest odd integer greater than $d_{fit}$ (denoted as $d$) to estimate the space-time overhead. The space-time overhead is $6d^3$ ($d^3$ unit cells) for the RHG cluster state and $d(2d-1)^2$ ($d$ rounds, $(2d-1)^2$ total qubits) for the XZZX surface code. We also directly calculated $d_{fit}^3$ to serve as an intuitive benchmark for comparison. }
    \label{tab1}
    
    \begin{tabular}{c|c|c|c|c|c|c} 
        \hline
        \multicolumn{1}{c}{Method} & 
        \multicolumn{1}{c}{Ansatz} & 
        \multicolumn{1}{c}{$p$} & 
        \multicolumn{1}{c}{$d_{fit}$} & 
        \multicolumn{1}{c}{$d$}&
        \multicolumn{1}{c}{Space-time Overhead} & 
        \multicolumn{1}{c}{$d_{fit}^3$} \\
        \hline
        
        \multirow{3}{*}{RHG-LD} & \multirow{3}{*}{$\ln p_L = -2.79 + 0.84 d \ln(p/p_{th})$} 
         &   $0.5\%$ & 24.34 & 25 & $6\times 25^3=93750$ & $14420.00$ \\
         & & $0.4\%$ & 20.56 & 21 & $6\times 21^3=55566$ & $8690.96$\\ 
         & & $0.3\%$ & 17.12 & 19 & $6\times 19^3=41154$ & $5018.11$\\ 
        \hline
        
        \multirow{3}{*}{XZZX-EC} & \multirow{3}{*}{$\ln p_L = -3.05 + 0.72 d \ln(p/p_{th})$} 
         &   $0.5\%$ & 21.90 & 23 & $23\times 45^2=46575$ & $10503.46$\\
         & & $0.4\%$ & 19.17 & 21 & $21\times 41^2=35301$ & $7044.87$\\
         & & $0.3\%$ & 16.52 & 17 & $17\times 33^2=18513$ & $4508.96$\\
        \hline
        
        \multirow{3}{*}{XZZX-BE} & \multirow{3}{*}{$\ln p_L = -3.29 + 0.66 d \ln(p/p_{th})$} 
         & $0.5\%$ & 20.42 & 21 & $21\times 41^2=35301$ & $8514.28$\\
         & & $0.4\%$ & 18.16 & 19 & $19\times 37^2=26011$ & $5988.35$\\
         & & $0.3\%$ & 15.89 & 17 & $17\times 33^2=18513$ & $4011.93$\\
        \hline
    \end{tabular}
\end{table}

Our results indicate that the RHG-LD method requires approximately $2\sim3$ times the space-time overhead compared to XZZX-EC or XZZX-BE. This additional overhead stems from two sources: First, at a same code distance, the RHG cluster state inherently incurs a space-time overhead factor of approximately $1.5$ relative to the XZZX surface code. Second, due to its slightly inferior logical performance, the RHG-LD method requires a marginally larger code distance to achieve the same level of error suppression. Notably, however, thanks to the advantage of RHG-LD in effective error distance $\beta$, the difference in the required code distance compared to XZZX-BE or XZZX-EC remains minimal. Here, we emphasize that while we utilize this ansatz to extrapolate logical error rates to larger code distances \cite{Baranes_2026}, it cannot be similarly applied to estimate performance at lower physical error rates, thereby claiming that a method with a larger effective error distance will ultimately yield a lower logical error rate. This is because, in the limit of vanishing physical error rates, the effective error distance for all methods is expected to converge to $d_e = (d+1)/2$ (or $\beta = 1/2$). Consequently, the current advantage in effective error distance will gradually diminish with decreasing physical error rates \cite{wu2022erasure}.\\

Furthermore, we use $d_{fit}^3$ as a direct proxy to approximate the overhead resulting from the second factor (the need for larger code distances). Compared to $d^3$, this metric avoids the rounding artifacts associated with restricting code distances to odd integers. We may calculate the ratio of $d_{fit}^3$ between RHG-LD and XZZX-BE or XZZX-EC with the same physical error rate. We observe that as the physical error rate decreases ($0.5\% \to 0.4\% \to 0.3\%$), the ratio of $d_{fit}^3$ between RHG-LD and XZZX-EC exhibits a progressively decreasing trend ($1.37 \to 1.23 \to 1.11$). This pattern is mirrored by the ratio between RHG-LD and XZZX-BE, which similarly drops from $1.69$ to $1.25$. This indicates that at lower physical error rates, the extra overhead of our RHG-LD method compared to the XZZX-BE/EC schemes is also reduced. This observation not only extends our main text's conclusion regarding the narrowing logical error rate gap at lower physical error rates, but also supplements the findings in Ref. \cite{Baranes_2026}, underscoring the necessity of higher-precision two-qubit gates (error rates $<0.5\%$).

\section{Benchmark and Discussion for Simulation Algorithms}\label{appF}
The primary metrics for evaluating quantum error correction simulation algorithms are accuracy (manifested as lower logical error rates or higher thresholds) \cite{PhysRevX.13.031007}  and time complexity (algorithm runtime) \cite{Gidney_2021,wu2023fusionblossomfastmwpm,Higgott2025sparseblossom}. Traditional QEC simulations typically comprise two components: (1) generating noisy circuits and sampling them to obtain syndrome and logical observable flips \cite{Gidney_2021}; and (2) utilizing the syndromes to decode (predict logical flips) and verifying these predictions against the ground truth \cite{wu2023fusionblossomfastmwpm,Higgott2025sparseblossom}. It is important to note that while accuracy is exclusively determined by the decoding algorithm (the latter component), time complexity benchmarks must account for the combined computational cost of both stages. In this section, we present runtime benchmarks for the algorithms employed in this work and discuss potential directions for future optimization.
\subsection{Benchmark for the Algorithm Runtime}
Before presenting the formal benchmarks, we analyze the individual components of the algorithm. Our approach utilizes \textit{Stim} to generate noisy circuit information  (encapsulated in a \textit{stim.DetectorErrorModel}) and construct the decoding graph \cite{Gidney_2021}, while \textit{PyMatching} is employed to perform the decoding \cite{Higgott2025sparseblossom}. When dealing with traditional Pauli errors, for a fixed set of parameters (including code distance and physical error rate), it suffices to generate a single noisy circuit and sample from it multiple times. In the case of erasure errors, however, erasures occurring at different locations correspond to distinct noisy circuits \cite{wu2022erasure}. Consequently, even with a fixed set of parameters, it is necessary to generate multiple noisy circuits and sample from each of them individually. Compared to standard approaches for handling erasure errors, our method incorporates an additional reweighting step. Specifically, this involves adjusting the edge weights of the decoding graph (representing the predicted error probabilities) based on the measurement results. In the standard treatment of erasure errors, since the error locations are precisely known, this operation can be automatically integrated into the conversion from the noisy circuit to the decoding graph. In our work, however, such information is not known a priori. We must instead utilize the final measurement results to infer potential error locations and assign corresponding probabilities to the resulting errors.\\

Each distinct leakage instance (referred to as circuit realization in \cite{PhysRevResearch.7.013249}) corresponds to a unique decoding graph. In practical implementation, however, generating separate noisy circuits and decoding graphs is relatively inefficient (a conclusion verified in the subsequent benchmarks). Therefore, we do not construct a new decoding graph for each sample \cite{PhysRevResearch.7.013249}. Instead, given a total requirement of $nshots$ samples, we generate only a fixed number of leakage instances ($num\_leakage\_instance$). For each instance, we construct a corresponding decoding graph and perform $nshots / num\_leakage\_instance$ samples on each graph. We find that a modest $num\_leakage\_instance$ produces logical error rates and variances indistinguishable from generating a unique graph per shot. We therefore fix $num\_leakage\_instance=16000$ in most simulations to optimize speed without compromising accuracy. In this benchmarking section, however, we adjust this parameter to evaluate the computational overhead of graph generation versus decoding independently. We define $\eta_{lea} := \frac{nshots}{num\_leakage\_instance}$ and set $\eta_{lea}=\frac{10^8}{16000}=6250$ as a typical case, and we also consider the runtime for $\eta_{lea} = \{10^6,1000,200,50\}$.\\

We prioritize the low-error-rate regime over the near-threshold regime because the resulting lower logical error rates necessitate a significantly larger number of samples to achieve statistically accurate results, which makes it the main cause of algorithm runtime. In our benchmark, all data in this section were obtained from simulations performed on a single core of a laptop equipped with an Intel Core i9-14900HX CPU, although our implementation supports efficient CPU parallelization (such as via \textit{joblib.Parallel}). The average runtime per shot is obtained by performing a total of $nshots=10^6$ simulations (with $num\_leakage\_instance$ set to $nshots/\eta_{lea}$) and dividing the total execution time by $nshots$. Our benchmark covers a wide range of odd code distances $d$, spanning from 7 to 21. We denote the runtime as $T$ and denote the runtime for decoding graph generation and the decoding process as $T_g$ and $T_d$, respectively. We denote $N$ as the number of nodes in the decoding graph \cite{Higgott2025sparseblossom}. In the RHG cluster state with code distance $d$, $N=d^3$.\\

\begin{figure*}[h]
\renewcommand{\thefigure}{S9}
    \centering
    \includegraphics[width=\textwidth]{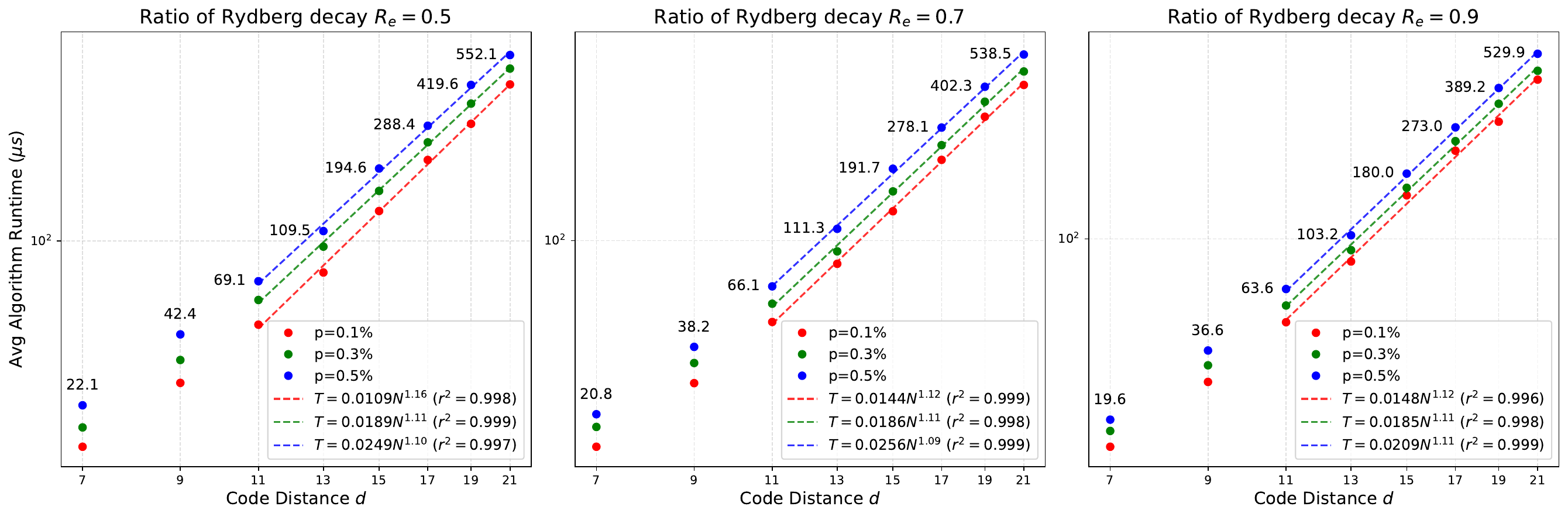}
    \caption{Typical case algorithm runtime and scaling. We perform the fitting over $d \in [11, 21]$ to exclude deviations arising from small-scale data. The specific runtime data for $p=0.5\%$ is annotated in black.}
    \label{figS9}
\end{figure*}

In the following, we present two sets of benchmarks. We first benchmark the typical algorithm runtime in this work, the average runtime per shot for RHG cluster states in the low physical error rate regime with $p=\{0.1\%,0.3\%,0.5\%\}$, $R_e=\{0.5,0.7,0.9\}$ and $\eta_{lea}=6250$. The results are shown in Fig.\ref{figS9}. These results demonstrate that our algorithm exhibits favorable scaling $T=O(N^{\beta}),\beta\approx 1.1$ in the typical case \cite{PhysRevLett.133.240602}, achieving a scaling comparable to traditional Pauli error decoding \cite{wu2023fusionblossomfastmwpm,Higgott2025sparseblossom}. We attribute the constant-factor time overhead compared to traditional Pauli decoding to the fact that our current algorithm possesses neither a high-speed sampler nor compatibility with the accelerated \textit{$Matching.decode\_batch()$} method in \textit{PyMatching}. We note the recent emergence of high-speed samplers capable of handling noisy circuits with erasure errors \cite{wu2025minimumweightparityfactordecoder}. Integrating such tools in future optimizations could further narrow the runtime gap between our algorithm and traditional Pauli error decoding. Furthermore, we observe that the algorithm's runtime decreases slightly as $R_e$ increases (see the runtime values for $p=0.5\%$ annotated in black in Fig. \ref{figS9}). In subsequent benchmarks, we confirm that the total runtime is dominated by the decoding process in typical cases. Consequently, this phenomenon indicates that decoding located errors possesses a slightly lower time complexity relative to traditional Pauli error decoding.\\

\begin{figure*}[h]
\renewcommand{\thefigure}{S10}
    \centering
    \includegraphics[width=0.5\textwidth]{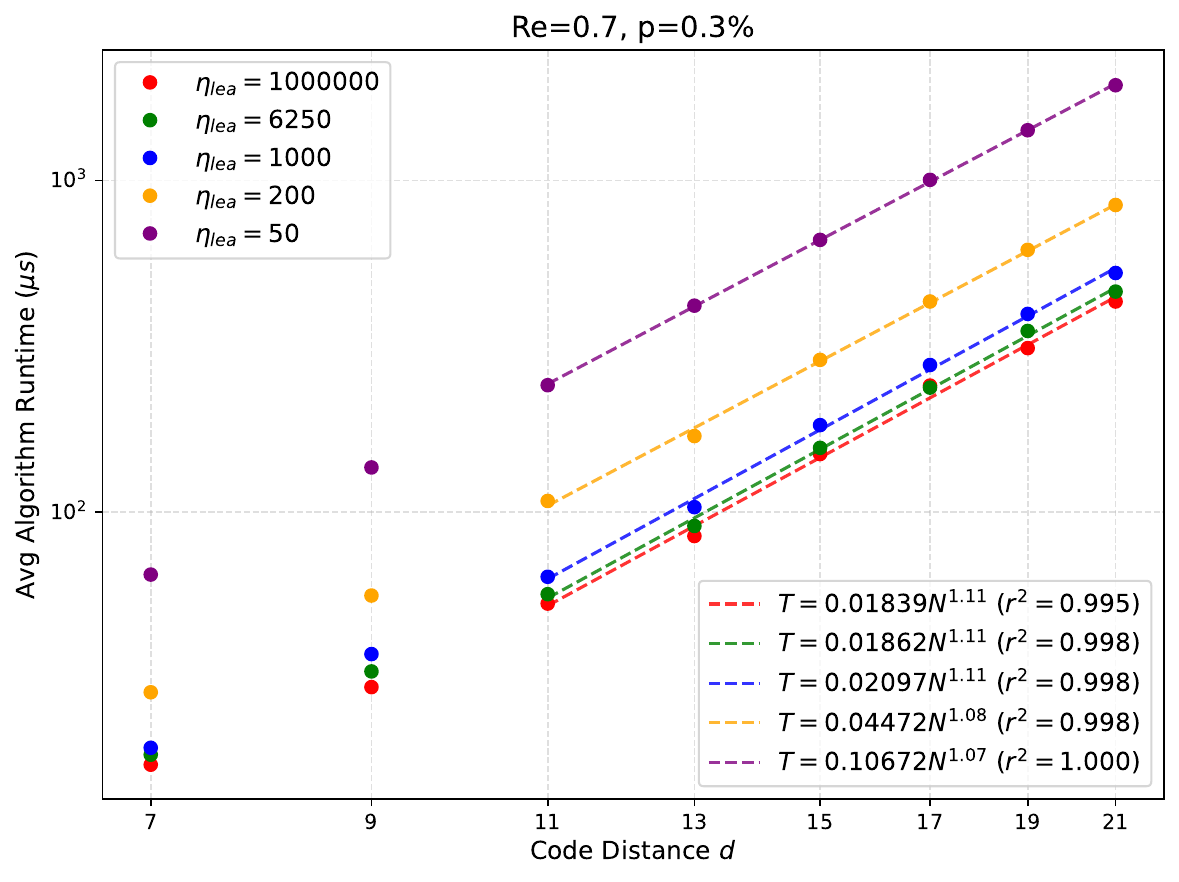}
    \caption{Algorithm runtime and scaling with a varing $\eta_{lea}$. We fixed $R_e=0.7$ and $p=0.3\%$ and other settings are similar to Fig.\ref{figS9}.}
    \label{figS10}
\end{figure*}

Next, we vary $\eta_{lea}\in\{10^6,6250(typical\;case),1000,200,50\}$ to investigate its impact on the decoding time ($T_d$) and the decoding graph generation time ($T_g$), as shown in Fig.\ref{figS10}. The total runtime $T$ and the runtime of the two independent components, $T_d$ and $T_g$, approximately satisfy the relationship $T \approx T_d + T_g / \eta_{lea}$. The results in Fig.\ref{figS10} reveal that constructing the decoding graph is approximately two orders of magnitude slower than the decoding algorithm itself. This confirms that at $\eta_{lea} = 6250$, the total runtime is dominated by the decoding algorithm rather than graph construction. Furthermore, the graph construction time follows the scaling relation $T_g = O(N^{\beta'})$ (where $\beta' < 1.1$), which aligns with our intuitive expectation that $T_g \approx O(N)$. These results further indicate that when simulations necessitate $\eta_{lea} < 200$ (such as when calculating logical error rates for $R_e > 0.95$), the construction of the decoding graph will become the dominant time overhead. In this future scenario, in addition to the efficient sampler and compatibility with faster $Matching.decode\_batch()$ methods mentioned above, a fast reweighting operation is also essential. Currently, we employ the $Matching.add\_edge()$ method in PyMatching to perform reweighting. We anticipate that the speed can be greatly improved by implementing the reweighting procedure directly on low-level C++ objects in PyMatching \cite{Gidney2024PyMatchingEdges}.

\subsection{Discussion on Future Optimizations for the Decoding Algorithm}
Having discussed potential avenues for reducing runtime, we now turn our attention to future improvements in terms of accuracy and utility. Regarding accuracy, we specifically focus on enhancing decoding performance for Rydberg decay or, more broadly, generalized leakage error models. We omit a detailed discussion on standard Pauli error optimization, as existing advancements in that area will naturally benefit our method when $R_e \neq 1$ \cite{PhysRevX.13.031007}.\\

\begin{figure*}[h]
\renewcommand{\thefigure}{S11}
    \centering
    \includegraphics[width=0.7\textwidth]{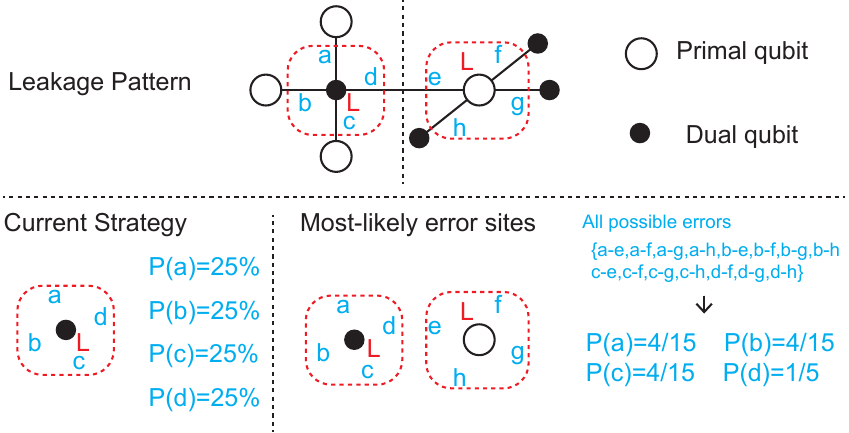}
    \caption{Two neighboring detected leakages and the gap between the current strategy and the most-likely error. Two neighboring leakage events are detected, with one on a primal qubit and the other on a dual qubit. The blue letters represent the error sites, which are identified as $L_i$ in Fig.2(b) of the main text. The set $\{a, b, c, d\}$ corresponds to error sites associated with leakage on the adjacent dual qubit, while $\{e, f, g, h\}$ corresponds to those associated with leakage on the adjacent primal qubit. Note that since the errors represented by sites $d$ and $e$ occur within the same CZ gate, they are mutually exclusive in our current error model (i.e., they cannot occur simultaneously). After accounting for all possible errors that could lead to these two leakage events, the occurrence probability of errors at the four sites adjacent to the dual qubit should be $P(a)=P(b)=P(c)=4/15$ and $P(d)=1/5$, as opposed to the $P(a)=P(b)=P(c)=P(d)=1/4$ estimated by our current strategy.
    }
    \label{figS11}
\end{figure*}

Decoding accuracy depends on the estimation of propagated error probabilities based on detected leakage information. The strategy employed in this work is an intuitive heuristic: we treat the primal and dual lattices separately. Specifically, when decoding the primal lattice, we utilize leakage information exclusively from dual qubits to infer the probability of error propagation. However, while this strategy roughly captures the relationship between error propagation probabilities and detected leakage, it does not fully resolve the problem of inferring the most-likely error from the leakage data \cite{Baranes_2026}. We present a simple example to illustrate the distinction between our current heuristic and the optimal inference of the most-likely error when two neighboring leakages are detected, as shown in Fig.\ref{figS11}. This example effectively demonstrates the general validity of the current heuristic approach, while also highlighting its limitations. Recent works have opted to solve this type of most-likely error problem directly; however, the associated computational complexity may hinder their scalability \cite{Baranes_2026}. Regarding the choice of algorithm to efficiently perform such probabilistic inference, we expect that integrating the Belief Propagation (BP) algorithm into existing matching-based decoders represents a promising future direction \cite{PhysRevX.13.031007}. Since the BP algorithm has been demonstrated to efficiently and accurately infer the posterior probabilities of errors based on observed syndromes, this scenario aligns closely with our objective of utilizing detected leakage information to infer the posterior probabilities of errors. Machine learning decoders also offer a compelling direction for future work \cite{Bluvstein_2025}. For such decoders, we assume the accuracy can be improved by employing neural architectures that incorporate physical priors and utilizing pre-training datasets that capture fine-grained error mechanisms \cite{Bluvstein_2025}.\\

For broader practical applicability, we envision extending our method to handle logical operations \cite{Baranes_2026}. Since logical gates in neutral atom platforms are typically realized via transversal gates, the logical operations necessitates the implementation of correlated decoding \cite{Bluvstein_2025,Zhou_2025,PhysRevLett.133.240602,cain2025fastcorrelateddecodingtransversal}. Recent works have demonstrated MWPM-based correlated decoding algorithms \cite{cain2025fastcorrelateddecodingtransversal}. Integrating our framework for handling Rydberg decay errors with these approaches could enhance logical performance while preserving scalability. We leave the specific implementation and detailed investigation to future work.

\section{Comparison with Several Relevant Recent Works}\label{appG}
In this section, we add a discussion between our work and two relevant recent works \cite{Baranes_2026,Perrin_2025}, focusing on atom loss, especially \cite{Baranes_2026} because they have considered \textit{teleportation-based QC}, which can be regarded as MBQC with a more compact circuit realization. For decoding the logical memory, teleportation-based QC and MBQC considered in our article are the same. Although all these works have shown high threshold and error distance $d_e \approx d$, our work differs from theirs in \textbf{key emphasis}, \textbf{error model}, \textbf{method to simulate Rydberg decay/atom loss}, and \textbf{decoder}.\\

\textbf{Key emphasis:} In our work, we have first developed this decoding strategy that allows final measurement, so we have attached importance to its \textit{basic properties}, including threshold and sub-threshold scaling. We have discussed the performance of the pure Rydberg decay error and the error model with the two-qubit depolarization error to reveal the general principle of this strategy by comparing with previous work on biased erasure \cite{sahay2023high}. In \cite{Baranes_2026}, they have attached more importance to a comprehensive computation framework of such a decoding strategy in logical algorithms. They have considered different types of logical circuits, including MBQC (\textit{teleportation-based QC} in their words) and \textit{modified Steane QC}. Since MBQC can be regarded as teleportation of logical state between code blocks in different time steps, their \textit{modified Steane QC} serves as a more flexible realization that implements multiple syndrome extraction in a single code block and then teleports the logical state to another code block when necessary. They also attach additional attention to biased error and combine this decoding strategy with correlated decoding \cite{PhysRevLett.133.240602}. \\

\textbf{Error model:} These works aim at different non-Pauli errors. Our work wants to deal with the \textit{Rydberg decay}, or more specifically, the leakage error from the Rydberg state during the two-qubit gate \cite{wu2022erasure,sahay2023high,ma2023high}, while their works \cite{Baranes_2026,Perrin_2025} want to deal with the \textit{atom loss}. There are two major differences between these two error models: 1. the Rydberg decay needs consideration of the coherent error \cite{sahay2023high,Bravyi_2018} while the general atom loss does not (See our \ref{appA} for detailed discussion); 2. the Rydberg decay needs consideration of the condition that a single leakage error dephases another qubit in two-qubit gate while general atom loss does not. The first difference lies in the fact that the Rydberg decay only occurs in atoms in the Rydberg state, which only couples with $\ket{1}$. This channel has off-diagonal terms in the process matrix, which can not be directly dealt with by the standard Pauli error model \cite{jayashankar2022achieving}. The second difference is that the Rydberg decay happens during the two-qubit gate, so a leakage during the gate time may dephase another atom \cite{wu2022erasure,sahay2023high}. Although atom loss can arise from conversion of BBR errors via anti-trapping, it also arises from another source - atom heating due to the coherent transport of atoms \cite{Baranes_2026,Perrin_2025,bluvstein2022quantum}. This error source doesn't need consideration of the two problems above, so their error model is reasonable from their perspective. \\

\textbf{Method to simulate Rydberg decay/atom loss:} Both these works have considered the simulation of atom loss by \textit{removing the gate} \cite{Baranes_2026,Perrin_2025} while our work considers it by regarding the Rydberg decay as an error. From a technical perspective, both these methods are reasonable, and we achieve comparable performance. From the analytical level, we argue that our method is better because when the Rydberg decay serves as an error, we can \textit{prove or analyse whether the distance is degraded}, as considered in previous works \cite{brown2020critical,PhysRevA.88.042308,PhysRevA.100.032325,d1v7-nctj,jandura2024surfacecodestabilizermeasurements}. This is helpful especially when the error model is complex, for example, when considering the correlated leakage \cite{zchg-x177}. Besides, we notice that \cite{Baranes_2026} uses a different method to deal with the leaked qubit. They use the \textit{supercheck} \cite{PhysRevLett.102.200501}, while we consider the leakage error as $50\%$ X and $50\%$ Z error \cite{wu2022erasure}. These two methods are equivalent according to the statement in \cite{PhysRevLett.102.200501}.\\

\textbf{Decoder:} Our work uses a matching-based decoder \cite{Higgott2025sparseblossom}. It has a polynomial complexity, and several efficient implementations exist \cite{Higgott2025sparseblossom,wu2023fusionblossomfastmwpm}. Instead, \cite{Baranes_2026} primarily uses a most-likely error decoder, which needs to handle hyperedge errors with an exponential worst-case complexity \cite{PhysRevLett.133.240602}  (although they have shown results with \textit{delayed-erasure decoder with MWPM}, such a decoder is only used to decode \textit{Free SE with period 0.25}, namely the erasure conversion \cite{wu2022erasure,sahay2023high}, as stated in its Supplementary Material). As benchmarked in \ref{appF}, our algorithm exhibits favorable runtime $T\sim N^{1.1}$ scaling. We have presented comparable thresholds in the logical memory with \cite{Baranes_2026}, but the polynomial complexity and the favorable runtime scaling enable scalability to large distances without fundamental difficulty.
\bibliography{supp_ref}